\documentclass[fleqn,usenatbib]{mnras}

\usepackage{newtxtext,newtxmath}

\usepackage[T1]{fontenc}

\DeclareRobustCommand{\VAN}[3]{#2}
\let\VANthebibliography\thebibliography
\def\thebibliography{\DeclareRobustCommand{\VAN}[3]{##3}\VANthebibliography}

\usepackage{graphicx}	%
\usepackage{amsmath}	%
\usepackage{MnSymbol}	%
\usepackage{cancel} 	%

\newcommand{\xHI}{\langle x_\mathrm{HI} \rangle}
\newcommand{\NHI}{N_\mathrm{HI}^\mathrm{DW}}
\newcommand{\Rb}{r_\mathrm{patch}}
\newcommand{\tQ}{t_\mathrm{Q}}
\newcommand{\logNHI}{\log\NHI}
\newcommand{\logNHIfull}{\log_{10} \NHI/\mathrm{cm}^{-2}}
\newcommand{\logtQ}{\log\tQ}
\newcommand{\logtQfull}{\log_{10} \tQ/\mathrm{yr}}

\title[Local quasar IGM damping wing constraints]{Inferring local quasar IGM damping wing constraints}

\author[Kist et al.]{
Timo Kist,$^{1}$\thanks{E-mail: kist@strw.leidenuniv.nl}
Joseph F. Hennawi$^{1,2}$
and Frederick B. Davies$^{3}$
\\
$^{1}$Leiden Observatory, Leiden University, P.O. Box 9513, 2300 RA Leiden,
The Netherlands\\
$^{2}$Department of Physics, University of California, Santa Barbara, CA 93106, USA\\
$^{3}$Max-Planck-Institut für Astronomie, Königstuhl 17, 69117 Heidelberg, Germany
}

\date{Accepted XXX. Received YYY; in original form ZZZ}

\pubyear{2025}

\begin{document}
\label{firstpage}
\pagerange{\pageref{firstpage}--\pageref{lastpage}}
\maketitle

\begin{abstract}
Lyman-$\alpha$ damping wings towards quasars are a highly sensitive probe of the neutral hydrogen (HI) content in the foreground intergalactic medium (IGM), not only constraining the global timing of reionization but also the \textit{local} ionization topology near the quasar. Near-optimal extraction of this information is possible with the help of two recently introduced reionization model-independent summary statistics of the HI distribution in the IGM \textit{before} the quasar started shining, complemented with the quasar's lifetime encoding the effect of its ionizing radiation as a third parameter. 
We introduce a fully Bayesian \texttt{JAX}-based Hamiltonian Monte Carlo (HMC) inference framework that allows us to jointly reconstruct the quasar's unknown continuum and constrain these local damping wing statistics. We put forward a probabilistic framework that allows us to tie these local constraints to any
specific reionization model and obtain model-dependent 
constraints on the global timing of reionization.
We demonstrate that we are able to constrain the (Lorentzian-weighted) HI column density in front of the quasar to a precision of $0.69_{-0.30}^{+0.06}\,\mathrm{dex}$ and its original distance to the first neutral patch before the quasar started shining
to $31.4_{-28.1}^{+10.7}\,\mathrm{cMpc}$ (if a noticeable damping wing is present in the spectrum), extracting hitherto unused local information from the IGM damping wing imprint. 
Once tied to a specific reionization model, we find that the statistical fidelity of our constraints on the global IGM neutral fraction and the lifetime of the quasar improves, while retaining the same precision as achieved by pipelines that infer these parameters directly.
\end{abstract}


\section{Introduction}

Little doubt exists about the fact that the formation of the first stars and galaxies heralded the epoch of reionization, a major phase transition in the history of our universe driven by the highly energetic radiation emitted by these objects which progressively reionized all neutral hydrogen (HI) in the intergalactic medium (IGM). Most details about this landmark event, however, are yet to be determined, starting with its mere timing in cosmic history, through to its topological features, informing us about the nature of the sources and sinks of reionization. One of the most promising probes of reionization is the Lyman-$\alpha$ transition in the spectra of bright astrophysical sources such as quasars. Its remarkable sensitivity to even the smallest amounts of neutral hydrogen in the IGM, causing extended Gunn-Peterson absorption troughs already at global volume-averaged IGM neutral fractions of $\xHI \gtrsim 10^{-4}$ \citep{gunn1965}, and the red damping wing induced by quantum mechanical line broadening when $\xHI$ becomes of order unity \citep{miralda-escude1998}, allow us to gain unique insights into how reionization proceeded over cosmic time.

While the first damping wing constraints have been reported for individual objects \citep{mortlock2011, bolton2011, greig2017b, greig2019, greig2022, banados2018, davies2018a, wang2020, yang2020, durovcikova2020, reiman2020}, \citet{greig2024b} and \citet{durovcikova2024} recently analyzed the first statistical ensembles probing the final stages of reionization. Even tighter constraints, reaching much deeper into the core stages of reionization, will be achievable in the coming years thanks to the growing number of high-redshift quasar spectra available to us \citep{dodorico2023, onorato2025, yang2025}, most prominently driven by the unprecedented discoveries of new objects at ever-higher redshifts in the Euclid wide field survey \citep{euclid_collaboration2019, banados2025, yang2025}.

Moreover, smooth roll-offs in the spectra near individual Gunn-Peterson troughs in the foreground of specific quasars have been interpreted as damping wings, possibly pointing to the
persistence of neutral islands in the IGM down to $5.5 \lesssim z \lesssim 6$ \citep{becker2024, spina2024, zhu2024, sawyer2025}. In addition to that, the advent of JWST has enabled the first damping wing constraints towards galaxies, pushing the highest-redshift frontiers \citep{curtis-lake2023, hsiao2024, keating2024b, park2025, umeda2024, umeda2025, mason2025}. However, additional care needs to be taken in this context due to the presence of intrinsic damped Lyman-$\alpha$ absorbers \citep[DLAs;][]{heintz2024, heintz2025} which can mimic the cosmological imprint from the IGM \citep{huberty2025}, introducing an additional nuisance process which needs to be marginalized out \citep{mason2025}.

Quasars, on the other hand, as powerful sources of ionizing radiation, suffer considerably less from this issue as they ionize away all residual neutral gas within their $\mathrm{Mpc}$-scale proximity zone. Any potentially remaining proximate DLA systems can be excluded based on the presence of associated weak metal absorption systems which can be identified with the help of supplementary high-quality 
spectra \citep{davies2025}. The remaining complication is the reconstruction of the intrinsic quasar continuum that the cosmological damping wing imprint has to be disentangled from. As the properties of these continua do not appear to evolve notably with redshift \citep{Shen2007}, data-driven models can be built based on unabsorbed low-redshift continua, and a plethora of approaches have been developed in the past years, ranging from simple models based on principal component analysis (PCA) to non-linear, neural network-based ones \citep[for a comprehensive overview, see][]{greig2024a}, most recently as a unified framework to \textit{jointly} reconstruct the quasar continuum and the IGM absorption imprint \citep{hennawi2025}.

\citet{kist2025a} demonstrated that approximately half the total error budget on the inferred IGM neutral fraction $\xHI$ is due to this continuum reconstruction task, whereas the other half is sourced by the stochastic distribution of IGM density fluctuations, as well as the distribution of neutral patches during reionization. To eliminate the latter part of this stochasticity from the inference task itself, \citet{kist2025b} introduced a new parameterization of IGM damping wings, probing the \textit{local} ionization topology in front of a given quasar \textit{before} it is altered by the quasar's ionizing radiation.
This three-parameter model comprises 1) the HI column density, weighted by a Lorentzian profile that accounts for the frequency dependence of the Lyman-$\alpha$ cross section, 2) the distance from the quasar to the first neutral patch, and 3) the quasar lifetime which encapsulates the effects of the quasar's ionizing radiation.

\citet{kist2025b} further demonstrated that these summary statistics are reionization model invariant,
and argued that they, when tied to a specific reionization model, provide constraints not only on the timing but also the topology of reionization. This work is concerned with the practical realization of this inference task by leveraging a fully Bayesian framework introduced in \citet{hennawi2025} which allows us to infer the first \textit{local} IGM damping wing constraints. Our local summary statistics can be constrained freely from any prior assumptions about the underlying reionization model. We demonstrate how the topology information from a given reionization model can be folded into these constraints \textit{subsequently} in a probabilistic manner, resulting in \textit{topology-informed} local constraints, and, in addition, a \textit{global} constraint on the IGM neutral fraction $\xHI$. In a companion paper, we apply this new approach to JWST/NIRSpec spectra of the two $z \sim 7.5$ quasars J1007+2115 and J1342+0928 \citep{kist2025c}.

We start by introducing this local parameter framework in Section~\ref{sec:theory}, and proceed in Section~\ref{sec:methods} by describing our inference approach originally introduced in \citet{hennawi2025} which is applicable both in the context of our new local IGM damping wing parameterization as well as the conventional global one, parameterized by the volume-averaged neutral fraction $\xHI$ and the lifetime $\tQ$ of the quasar. In Section~\ref{sec:inference_results} we test the statistical fidelity of the pipeline and quantify the precision of the local damping wing constraints. We also compare the precision of the resulting constraints on $\xHI$ and $\tQ$ to that of the directly inferred ones. We conclude in Section~\ref{sec:conclusions}. %

\section{Theory: relating a local IGM damping wing model to the global timing of reionization}
\label{sec:theory}

Due to its sensitivity to neutral hydrogen in the foreground IGM, the Lyman-$\alpha$ damping wing signature in the spectra of high-redshift sources is considered one of the key probes of the global volume-averaged IGM neutral fraction $\xHI$ as a function of redshift. As has recently been noted \citep{chen2024, keating2024a, kist2025b}, considerable sightline-to-sightline variations are possible due to fluctuations in the cosmological density field as well as the patchy nature of reionization. Specifically, \citet{kist2025b} identified a two-dimensional set of physical summary statistics that (along with the lifetime of the quasar as a third parameter) tightly parametrizes the characteristic shape of the IGM damping wing. These two statistics are informative not only about the global timing of reionization, but also the local ionization topology \textit{before the quasar started shining}. Note that the quasar radiation inevitably modifies the topology that ultimately imprints the damping wing, but our physical interest is directed at the original topology which encapsulates the information about what we will henceforth refer to as the \textit{pre-quasar} IGM, in contrast to the \textit{post-quasar} one, impacted by the quasar's ionizing radiation.
In this section, we will provide a short but self-contained overview over this parameterization, and demonstrate subsequently how we can fold in the topological information that arises from the assumed reionization model through a prior on our local damping wing statistics, and how we can use this to relate back these summaries to the global timing of reionization, parameterized by the redshift evolution of the global IGM neutral fraction $\xHI(z)$.

\subsection{A local, topology-independent parameterization of quasar IGM damping wings}
\label{sec:labels}

We start by defining the two summary statistics of the local ionization topology around quasars introduced in \citet{kist2025b}. We refer the reader to this work for all additional details. The authors demonstrated that the information contained in the damping wing optical depth $\tau_\mathrm{DW}$ can largely be condensed into a single number, resulting in a near-optimal parameterization of the Lyman-$\alpha$ damping wing across the entire spectral range. For a quasar at redshift $z_\mathrm{QSO}$ with a post-quasar HI density field $n_\mathrm{HI}^\mathrm{post}$, 
the damping wing optical depth $\tau_\mathrm{DW}$ as a function of Lyman-$\alpha$ rest-frame wavelength $\lambda_\mathrm{rest} = \lambda_\alpha (1+\frac{v}{c})$ is given by
\begin{equation}
\label{eq:tau_DW}
    \tau_\mathrm{DW}(\lambda_\mathrm{rest}) = \int_{0}^{R(z_\mathrm{QSO})} n_\mathrm{HI}^\mathrm{post}(R) \cdot \sigma_\alpha\left(\frac{1+z_\mathrm{QSO}}{1+z(R)}\,\lambda_\mathrm{rest}\right) \; \mathrm{d}R,
\end{equation}
where $\sigma_\alpha$ is the Lyman-$\alpha$ cross section, $\mathrm{d}R$ the infinitesimal proper line-of-sight interval, and $R(z_\mathrm{QSO})$ the corresponding proper distance from the observer to the quasar.

In essence, Eq.~(\ref{eq:tau_DW}) constitutes a column density integral of the HI density field with an additional weighting kernel, governed by the Lyman-$\alpha$ cross section $\sigma_\alpha$. To excellent approximation, $\sigma_\alpha$ at a given spectral velocity pixel $v_\mathrm{T}$ 
is of Lorentzian shape, i.e., $\sigma_\alpha \sim (v-v_\mathrm{T})^{-2}$, whose impact we can capture by defining the Lorentzian line-of-sight average $\llangle . \rrangle_{\mathrm{Lor}}$ 
of a field $X$ (such as the HI density field $n_\mathrm{HI}$) as
\begin{equation}
\label{eq:Lorentzian_avg}
    \llangle X \rrangle_{\mathrm{Lor}} \equiv \frac{1}{\mathcal{N}(u_\mathrm{min}, u_\mathrm{max})} \, \int_{u_\mathrm{min}}^{u_\mathrm{max}}  \frac{X(u)}{(u + 1)^2} \; \mathrm{d}u,
\end{equation}
where $u=R/R_\mathrm{T}$ is a dimensionless integration variable, and $\mathcal{N}(u_\mathrm{min}, u_\mathrm{max}) \equiv (u_\mathrm{max}-u_\mathrm{min})/((u_\mathrm{max}+1)(u_\mathrm{min}+1))$ a normalization factor ensuring $\llangle \boldsymbol{1} \rrangle_{\mathrm{Lor}} = 1$ for the identity field $\boldsymbol{1}$. Here we defined $R_\mathrm{T}$ as the (positive) proper
distance value corresponding to the red-side velocity offset $v_\mathrm{T}$ via $R_\mathrm{T} \equiv +v_\mathrm{T}/H(z_\mathrm{QSO})$.

Instead of directly taking the Lorentzian-weighted 
average of the post-quasar HI density field $n_\mathrm{HI}^\mathrm{post}$, we operate on its pre-quasar version $n_\mathrm{HI}^\mathrm{pre} = \langle n_{\mathrm{H}} \rangle(z_\mathrm{QSO}) \cdot x_\mathrm{HI} \cdot \Delta$, unaffected by the ionizing quasar radiation and therefore directly informative about the pre-quasar ionization topology parameterized by the neutral fraction field $x_\mathrm{HI}$, along with the dimensionless matter overdensity field $\Delta$ and the cosmic mean hydrogen density $\langle n_{\mathrm{H}} \rangle(z_\mathrm{QSO})$.
Specifically, we define the \textit{Lorentzian-weighted} 
HI column density
\begin{align}
\label{eq:NHI_DW}
    N_\mathrm{HI}^\mathrm{DW} \equiv\; &5.1 \times 10^{20}\,\mathrm{cm}^{-2}\times \left(\frac{\mathcal{N}\left(\tfrac{r_\mathrm{min}}{r_\mathrm{T}}, \tfrac{r_\mathrm{max}}{r_\mathrm{T}}\right)}{0.67}\right) \left(\frac{r_\mathrm{T}}{18\,\mathrm{cMpc}}\right) \\ \nonumber
    &\times \left(\frac{1+z_\mathrm{QSO}}{1+7.54}\right)^2 \left(\frac{\llangle x_\mathrm{HI} \cdot \Delta \rrangle_{\mathrm{Lor}}}{1}\right),
\end{align}
where the normalization factor is defined as above, and $r_\mathrm{min}$, $r_\mathrm{max}$ and $r_\mathrm{T}$ are the comoving versions of the proper distances $R_\mathrm{min}$, $R_\mathrm{max}$ and $R_\mathrm{T}$.\footnote{Note that this is the convention we will adopt for all distances throughout this work.} With this definition at hand, it is then straightforward to show that we have identified an asymptotically optimal parameterization of the (pre-quasar) IGM damping wing, in the sense that at the velocity offset $v_\mathrm{T}$ itself, we find a direct proportionality between the optical depth and our summary statistic:
\begin{equation}
\label{eq:tau_DW_approx}
    \tau_\mathrm{DW}^\mathrm{pre}(v=v_\mathrm{T}) \simeq \frac{e^2}{m_e c^2}\,f_\alpha\,\gamma_\alpha\,\lambda_\alpha\,(c/v_\mathrm{T}-1)^2 \times \NHI
\end{equation}
in the limit where the integration limits approach $u_\mathrm{min} \to 0$ and $u_\mathrm{max} \to R(z_\mathrm{QSO}) / R_\mathrm{T}$, and where the Lorentzian approximation of the Lyman-$\alpha$ cross section is valid.

To make the statistic adequate for the \textit{post}-quasar optical depth despite the fact that we are operating on the \textit{pre}-quasar HI density field, we have to account for the fact that quasars commonly ionize away all neutral material within the first few $\mathrm{cMpc}$ surrounding them, and therefore not contributing to the optical depth integral in Eq.~(\ref{eq:tau_DW}). We do so by starting the integration in Eq.~(\ref{eq:Lorentzian_avg}) at a common size of this ionized bubble, $r_\mathrm{min} = 4\,\mathrm{cMpc}$. %
We fix the upper integration limit to $r_\mathrm{max} = r_\mathrm{min} + 100\,\mathrm{cMpc}$ since any neutral patches beyond this distance are down-weighted to such a high degree by the Lorentzian weighting kernel that they do not notably contribute to $\NHI$ anymore. Lastly, we set the reference distance $r_\mathrm{T}$ with respect to which our summary statistic is defined, to $r_\mathrm{T} = 18\,\mathrm{cMpc}$ (corresponding to $v_\mathrm{T} \simeq 2000\,\mathrm{km}/\mathrm{s}$ at $z_\mathrm{QSO} = 7.54$), noting that we are not sensitive to this choice since the damping wing largely constitutes a one-parameter family, i.e., the imprint is so correlated that its value at one spectral pixel $v_\mathrm{T}$ closely determines its value at all other spectral pixels,
as demonstrated in \citet{kist2025b}. As the parameter range for $\NHI$ spans several orders of magnitude, we are in fact sensitive to the logarithmic quantity $\logNHIfull$, and for notational simplicity we henceforth adopt $\logNHI$ as short-hand notation for this.

We adopt as a second summary statistic the distance $\Rb$ between the source and the first neutral patch in the \textit{pre}-quasar topology. This quantity has been constrained by \citet{mason2025} 
in the context of galaxy IGM damping wings, and has been shown in \citet{kist2025b} to remain a meaningful summary for \textit{quasar} IGM damping wings---despite the impact of the ionizing quasar radiation due to which the pre- and post-quasar distance to the first neutral patch do not necessarily agree \citep[c.f.][who inferred the post-quasar version of this distance]{schroeder2013}. Most importantly, $\Rb$ encapsulates information complementary to $\logNHI$ as demonstrated in \citet{kist2025b}. %

\subsection{Folding in the topology dependence and constraining the global IGM neutral fraction}
\label{sec:top_dep}

\citet{kist2025b} showed that the two local summary statistics $\logNHI$ and $\Rb$ 
not only minimize the scatter of IGM transmission values (i.e., the amount of information that these statistics do not explain)
in the damping wing region of the spectrum to $\lesssim 1\,\%$ across the entire range of physical parameter space, but that these two summaries are largely insensitive to the underlying reionization topology in the sense that the median and the $68$-percentile scatter of the IGM transmission profiles at a given set of parameter values are identical, regardless of what reionization topology the underlying sightlines originate from. This implies that the specific distribution of neutral patches along a given sightline does not impact the damping wing shape, provided the $\logNHI$ and $\Rb$ parameter values are fixed. All topological information is instead fully encoded in the statistical distribution of these statistics \textit{within} a given topology. As a result, in a Bayesian sense, all assumptions about the ionization topology can be absorbed into the prior distribution imposed on $(\logNHI, \Rb)$, while the IGM transmission likelihood, given $\logNHI$ and $\Rb$ as model parameters, remains reionization model independent and can be 
determined by considering models of damping wing transmission profiles from \textit{any} fiducial reference topology. We exploit this fact by constructing the likelihood based on sightlines generated according to the simplistic toy prescription introduced in \citet{kist2025b} which augments the smoothness of the likelihood as a function of $\logNHI$ and $\Rb$, facilitating the practical task of sampling from the posterior distribution. The realistic topology only enters the analysis for determining the prior, and for converting the \textit{local} $(\logNHI, \Rb)$ constraints into a \textit{global} $\xHI$ constraint. We do so by probabilistically relating global and local parameters to one another based on how the local parameters are distributed in a given global topology.

\subsubsection{Simulating IGM transmission profiles}
\label{sec:sims}

\begin{figure}
	\includegraphics[width=\columnwidth]{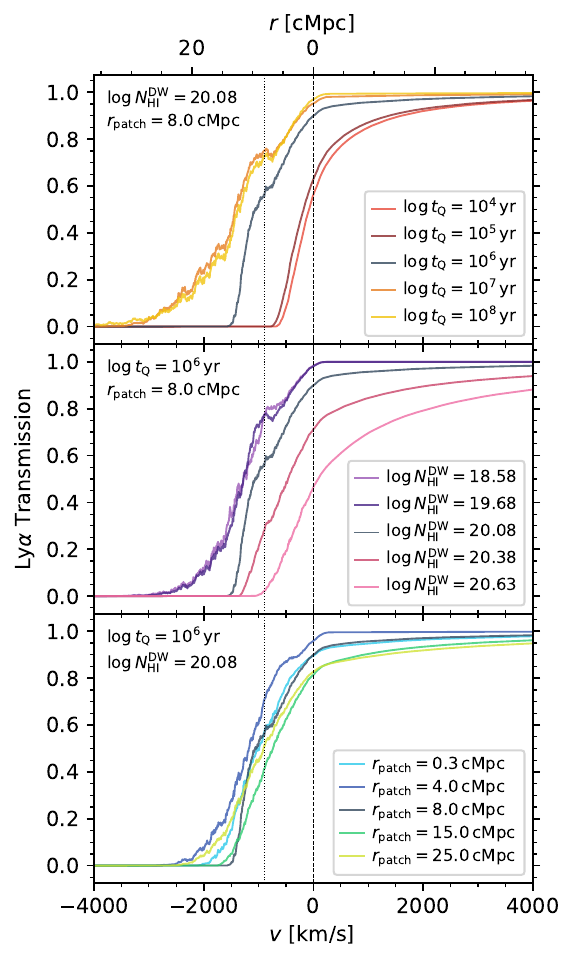}
    \caption{Median IGM transmission profiles in our local damping wing parameterization. Variations with respect to $\tQ$, $\logNHI$ and $\Rb$ are displayed in the upper, middle and lower panel, respectively. When fixed, these parameters are set to $\tQ = 10^6\,\mathrm{yr}$, $\logNHI = 20.08$ and $\Rb = 8\,\mathrm{cMpc}$ (shown as black reference line in each panel). The dotted vertical line marks the spectral location corresponding to the reference choice of $\Rb$. All profiles are simulated following the prescription in Section~\ref{sec:sims} with synthetic $x_\mathrm{HI}$ profiles.}
    \label{fig:transm_median}
\end{figure}

We simulate IGM transmission profiles following \citet{davies2018a}'s hybrid approach of combining sightlines from cosmological hydrodynamical simulations with independent neutral fraction skewers and 1d radiative transfer. We extract $3600$ density, velocity and temperature skewers originating at the $600$ most massive halos 
with masses of $M_\mathrm{halo} \geq 1.3\times 10^{11}\,M_\odot$,
and extending towards the six principal directions of the $z = 7.0$ snapshot of the Nyx hydrodynamical simulations \citep{almgren2013, lukic2015}. The simulation box measures $100\,\mathrm{cMpc}/h$ on a side and contains $4096^3$ baryon and dark matter particles, respectively. 

We simulate neutral fraction $x_\mathrm{HI}$
skewers in two different manners, 1) by extracting them from realistic semi-numerical reionization topologies, used to determine our priors and convert our local constraints to a \textit{global} $\xHI$ constraint on the timing of reionization, or 2) by producing synthetic $x_\mathrm{HI}$ skewers according to a simple toy prescription which we use to determine the IGM transmission likelihood.

The former skewers originate from topologies generated using a modified version of the \texttt{21cmFAST} code \citep{mesinger2011, davies2022} at fixed global IGM neutral fractions of $\xHI = 0.0, 0.05, ..., 1.0$. Intermediate values between zero and one are achieved by tuning the ionizing efficiency $\zeta$. From each simulation box (of size $400\,\mathrm{cMpc}$ on a $2048^3$ initial and $512^3$ output grid), we extract $20$ randomly oriented skewers originating from each of the $500$ most massive halos of masses $M_\mathrm{halo} \geq 3\times 10^{11}\,M_\odot$, giving a total of $10,000$ $x_\mathrm{HI}$ skewers.
We then combine each of our $3600$ 
hydrodynamical Nyx sightlines with a random draw from these $10,000$ 
neutral fraction skewers. By construction, these skewers vary smoothly as a function of the global IGM neutral fraction $\xHI$. We can also measure the local summary statistics $\logNHI$ and $\Rb$ for each individual sightline,\footnote{Note that before determining $\Rb$, we smooth the $x_\mathrm{HI}$ field with a box-car filter of size $0.5\,\mathrm{cMpc}$ and define $\Rb$ as the distance from the quasar where this smoothed field first exceeds a neutral fraction threshold of $50\,\%$ to avoid being overly sensitive to extremely small-scale fluctuations in the $x_\mathrm{HI}$ field.} and aggregate all sightlines according to these labels; however, this does not guarantee that these aggregated profiles vary smoothly as a function of $\logNHI$ and $\Rb$ since the number of available sightlines can vary significantly in different regions of $(\logNHI, \Rb)$ parameter space.

Such small sample sizes have a particularly unfavorable impact on the noise level of the covariance matrices we have to estimate for determining the likelihood function (see Eq.~(\ref{eq:likelihood})). Noisy covariance matrices can introduce discontinuities to this, and sampling from such a non-smooth distribution becomes practically impossible.
We can avoid such issues by instead generating synthetic $x_\mathrm{HI}$ skewers according to the toy prescription introduced in \citet{kist2025b} which by construction ensures smooth variations of the resulting skewers with respect to these local summaries. In short, we start from a given density sightline in its completely ionized state, and subsequently add in neutral patches of minimum size $\Delta r_\mathrm{min} = 0.5\,\mathrm{cMpc}$, growing them out one-by-one to a maximum size $\Delta r_\mathrm{max} = 5.0\,\mathrm{cMpc}$ until a desired HI column density $\logNHI$ is reached. By definition, the position of the first neutral patch is set by $\Rb$, and the subsequent patches are added according to a fixed sequence of neutral patch locations for a given density sightline, ensuring continuity of the resulting skewers with respect to $\logNHI$. We emphasize that achieving continuity 
is the sole purpose of this particular approach, and that it does not seek to model any realistic physical processes occurring during reionization---in fact, \citet{kist2025b} demonstrated that a physical reionization model 
is not necessary for our purposes of determining the IGM transmission likelihood, as the mean and variance of the resulting IGM transmission profiles in our local damping wing parameterization only depend on the distribution of density fluctuations in the IGM but \textit{not} on the specific distribution of neutral patches along the line of sight.

Further, it is important to note that the minimum and maximum column density values $\logNHI$ are sightline-dependent, set by the distribution of IGM density fluctuations along the line of sight for a completely neutral and a completely ionized sightline, respectively. At a given location in $(\logNHI, \Rb)$ parameter space, we can therefore end up with a number between $0$ and $3600$ sightlines, depending on how many sightlines allow for this specific parameter combination. We compute these synthetic $x_\mathrm{HI}$ skewers on an irregular $(\logNHI, \Rb)$ parameter grid, reflecting the degree to which these parameters impact the resulting IGM transmission profiles. 
Specifically, our parameter grid consists of $21$ column density values between $17.48 \leq \logNHI \leq 21.08$, and $18$ neutral patch distances between $0.3\,\mathrm{cMpc} \leq \Rb \leq 143.0\,\mathrm{cMpc}$. A finer grid spacing is chosen towards higher column densities $\logNHI$ and shorter neutral patch distances $\Rb$ where small parameter variations have a noticeable impact on the damping wing strength, whereas this is not necessary at low $\logNHI$ and high $\Rb$ where no damping wing is present.

Finally, we combine hydrodynamical and neutral fraction skewers, adopting the Nyx temperature field for all ionized regions, and assuming an initially cold IGM with $T = 2000\,\mathrm{K}$ for all neutral patches. We then perform one-dimensional radiative transfer along these sightlines to model the impact of the ionizing quasar radiation following \citet{davies2016}. Our model resembles the $z_\mathrm{QSO} = 7.54$ quasar ULAS J1342+0928 with a simple light bulb light curve corresponding to an ionizing photon emission rate of $Q = 10^{57.14}\,\mathrm{s}^{-1}$ for quasar lifetimes $\tQ$ on a logarithmic grid with $51$ values between $\tQ = 10^3$ and $10^8\,\mathrm{yr}$, and we henceforth adopt $\logtQ$ as short-hand notation for $\logtQfull$. To produce Lyman-$\alpha$ transmission profiles, we finally convolve the physical output fields with a Voigt profile \citep{tepper-garcia2006}.

As there is no star formation prescription in the Nyx simulations, strong proximate optically thick absorption line systems are not modeled realistically. This means we have to remove such sightlines from our full set of simulated IGM transmission profiles. This does not pose a major limitation to our approach as we do not aim to constrain the ionization state of the IGM based on targets where such strong absorption systems are present. As their absorption signature could easily be confused with the intergalactic one, this would introduce a significant additional modeling uncertainty when trying to disentangle it from the IGM damping wing, similar to what is required for galaxies \citep{mason2025}.
Observationally, such sightlines can be excluded a priori by identifying associated metal absorption lines in the spectrum of the source \citep{davies2025}. In our simulated profiles, we identify such sightlines by computing the HI column density within chunks of size $0.1\,\mathrm{pMpc}$ along the fully ionized realization of each sightline, and excluding all those sightlines containing at least one chunk with an (unweighted) HI column density of at least $10^{19}\,\mathrm{cm}^{-2}$ within the first $5000\,\mathrm{km}/\mathrm{s}$ from the source, leaving $2545$ 
of the total of $3600$ 
sightlines that can be used for the subsequent analysis.

We depict in Figure~\ref{fig:transm_median} the median profiles based on the synthetic $x_\mathrm{HI}$ skewers as a function of the three parameters of our local parameterization. From top to bottom, we show the variation with respect to $\logtQ$, $\logNHI$ and $\Rb$, respectively. The parameters which are not varied in a given panel are fixed to the reference values of $\tQ = 10^6\,\mathrm{yr}$, $\logNHI = 20.08$ and $\Rb = 8\,\mathrm{cMpc}$ (black line in each panel). We observe the well-known trend of larger proximity zones for long-lived quasars due to the increasing size of the ionized bubble such objects have carved out around themselves \citep[see e.g.][]{chen2021}. In line with this, the strength of the IGM damping wing decreases with increasing $\logtQ$ due to the decreasing amount of neutral hydrogen along the line of sight. Similarly, the damping wing strength increases and the proximity zone size decreases with increasing HI column density $\logNHI$. Note, however, that this parameter captures the properties of the \textit{pre}-quasar ionization topology, whereas $\tQ$ encapsulates the effects of the quasar's ionizing radiation.

The functional dependence of the profiles on the distance $\Rb$ from the source to the first neutral patch is more complex. First, note that the median profiles in the two upper panels show a clear bump in transmission at $\Rb = 8\,\mathrm{cMpc}$. This is because unlike the already ionized material closer to the quasar, this initially neutral patch at $\Rb = 8\,\mathrm{cMpc}$ receives an additional amount of photoelectric heating when ionized by the quasar, leading to an enhancement in transmission in the corresponding region of the spectrum \citep[c.f.][]{kist2025a}. We can see in the bottom panel of Figure~\ref{fig:transm_median} that this position shifts with the distance the first neutral patch, with the $\Rb = 4\,\mathrm{cMpc}$ profile showing a clear bump at that location instead. The remaining shape of the profiles depends highly non-trivially on the value of $\Rb$. For values of $0.3\,\mathrm{cMpc} \leq \Rb < 4\,\mathrm{cMpc}$, the first neutral patch is located \textit{outside} the integration range of $\logNHI$ which starts at $r_\mathrm{min} = 4\,\mathrm{cMpc}$. As a result, $\Rb$ is the only statistic capturing information about any pre-quasar neutral material in this region, and hence acts as a summary akin to $\logNHI$, with increasingly strong absorption for smaller values of $\Rb$.

This behavior turns around for $\Rb \geq 4\,\mathrm{cMpc}$, where the damping wing shape is already captured to lowest order by $\logNHI$. Contrary to the case where $\Rb < 4\,\mathrm{cMpc}$, we find a mild increase in the strength of the absorption signature with increasing $\Rb$. This is because we are concerned with \textit{pre}-quasar statistics, and at fixed HI column density $\logNHI$, a neutral patch is more likely to be ionized away by the radiation of a quasar of a given lifetime $\logtQ$ the closer it is located to the source, i.e., the smaller the value of $\Rb$. However, more nearby neutral patches are also the greatest contributors to the damping wing optical depth, and as such, the damping wing strength is lower the closer the first neutral patch was located to quasar, i.e., the more likely it was to get ionized away. When comparing to the upper two panels, however, we see that these effects are clearly subdominant to the impact of quasar lifetime $\logtQ$ and HI column density $\logNHI$ on the IGM damping wing strength. However, the fact that the transmission profiles still change as a function of $\Rb$ when $\logNHI$ and $\logtQ$ are fixed shows that this parameter carries information complementary to that encapsulated by the other two parameters.

\subsubsection{The prior on the local summary statistics induced by the reionization model}
\label{sec:prior}

\begin{figure*}
	\includegraphics[width=\textwidth]{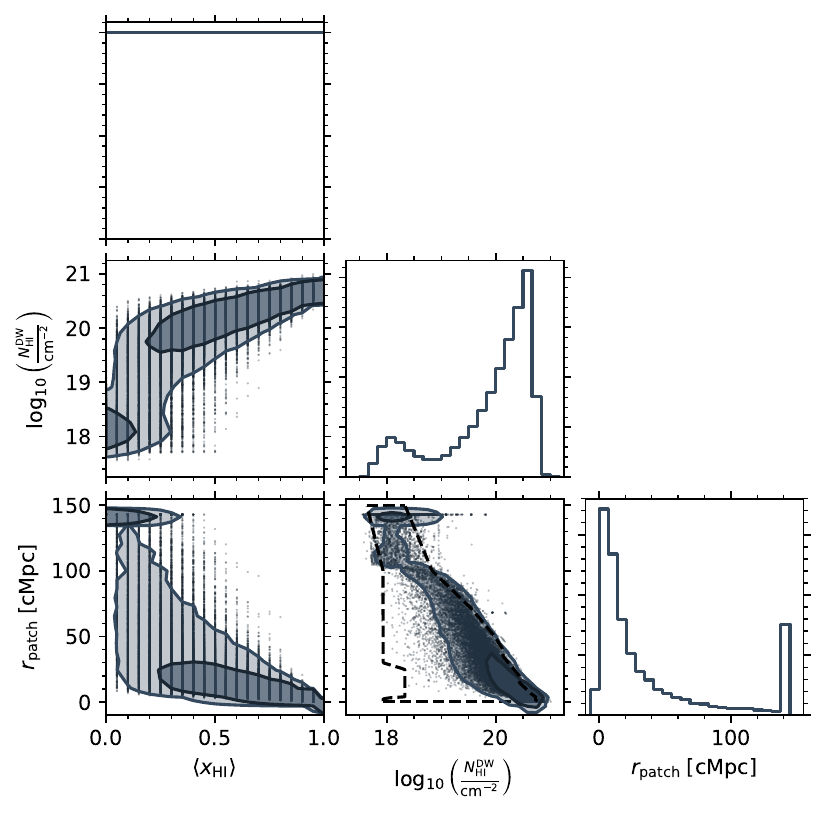}
    \caption{Joint distribution $P(\xHI, \logNHI, \Rb)$ of the global IGM neutral fraction $\xHI$ and the local summary statistics $\logNHI$ and $\Rb$, determined based on $21 \times 2545$ sightlines from the realistic semi-numerical reionization topology introduced in Section~\ref{sec:sims}. Black dots denote individual sightline data which is arranged on a grid in the $\xHI$ dimension since the sightlines were extracted from $21$ distinct topologies with fixed global IGM neutral fractions between $0 \leq \xHI \leq 1$. Contours correspond to the $68\,\%$ and $68\,\%$ percentile regions. The dashed line in the $(\logNHI, \Rb)$ plane encloses the physically permitted domain for these parameters, determined based on simulations run with the synthetic prescription described in Section~\ref{sec:sims}. Note that this boundary is constructed based on where we have at least $800$ sightlines available to estimate means and covariances, and hence individual, rare realizations can lie outside of it.}
    \label{fig:P_xHI_NHI_Rb}
\end{figure*}

\begin{figure*}
    \includegraphics[width=0.75\textwidth]{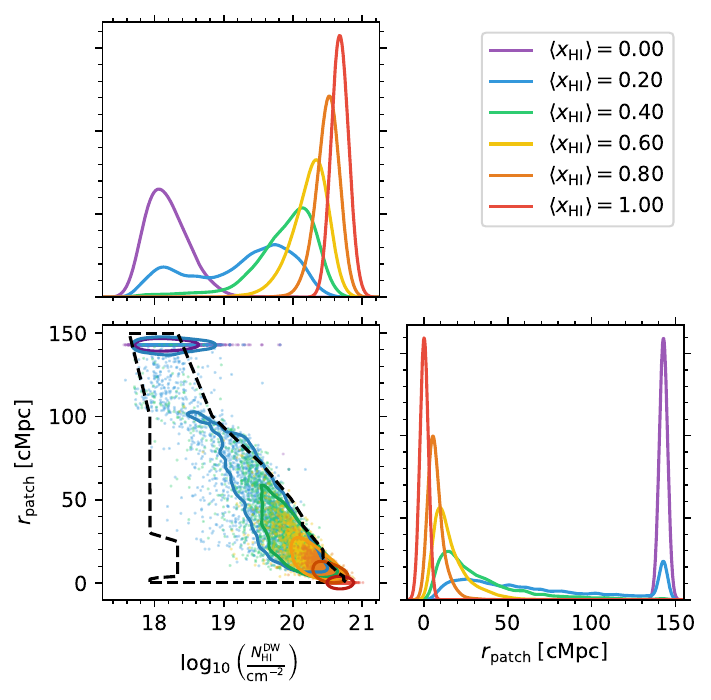}
    \caption{Conditional distribution $P(\logNHI, \Rb | \xHI)$ of the two local summary statistics $\logNHI$ and $\Rb$ given the global IGM neutral fraction $\xHI$, determined based on $2545$ sightlines from the realistic semi-numerical reionization topology introduced in Section~\ref{sec:sims} at six different global IGM neutral fractions between $0 \leq \xHI \leq 1$. Dots denote individual sightline data, and contours the $68\,\%$ scatter. For visibility, we smoothed the one-dimensional marginal distributions with a kernel density estimation (KDE). The dashed line in the $(\logNHI, \Rb)$ plane encloses the physically permitted domain for these parameters, determined based on the simulations run with the synthetic prescription described in Section~\ref{sec:sims}.}
    \label{fig:P_NHI_Rb_given_xHI}
\end{figure*}

As demonstrated in \citet{kist2025b}, the mean and variance
of the IGM transmission profiles at fixed $(\logNHI, \Rb)$ parameter values do not depend on the reionization topology that the sightlines originate from. Instead, the topology exclusively determines \textit{how common} a given parameter combination is. In other words, a statistical distribution on these parameters---be it prior or posterior---can directly be translated between the global $\xHI$ parameterization and the local $(\logNHI, \Rb)$ parameterization.

When constraining the timing of reionization with quasar IGM damping wings, we do not want to impose any prior assumptions on the reionization state of the universe, i.e., a priori, we consider any global IGM neutral fraction value $\xHI$ between $0$ and $1$ as equally likely. Formally speaking, we impose a flat prior $P(\xHI) = \mathrm{Unif}_{[0, 1]}(\xHI)$ on the global IGM neutral fraction, where $\mathrm{Unif}_{[a, b]}(x)$ is a uniform distribution of the random variable $x$ on the interval $[a, b]$.

But the assumed reionization topology determines the conditional distribution $P_\mathrm{top}(\logNHI, \Rb \,|\, \xHI)$ of the local summaries given $\xHI$, since clearly different local parameter configurations are differently common in different topologies with different global IGM neutral fractions, and henceforth we denote all distributions $P_\mathrm{top}$ that are affected by these topology assumptions with a corresponding subscript. Based on this, we can use the global prior $P(\xHI)$ to write down the joint prior distribution on all three parameters:
\begin{align}
\label{eq:joint}
    P_\mathrm{top}(&\xHI, \logNHI, \Rb) \\
    &= P_\mathrm{top}(\logNHI, \Rb \,|\, \xHI) \times P(\xHI). \nonumber
\end{align}
Marginalizing Eq.~(\ref{eq:joint}) over the global IGM neutral fraction, we see that the simple uniform prior on $\xHI$ translates into a non-trivial, topology-informed prior on the local parameters $(\logNHI, \Rb)$, given by
\begin{align}
\label{eq:prior}
    P_\mathrm{top}(&\logNHI, \Rb) \\
    &= \int\mathrm{d}\xHI \; P_\mathrm{top}(\logNHI, \Rb \,|\, \xHI) \times P(\xHI). \nonumber
\end{align}
It is this step where the assumptions about the reionization model affect our constraints: different reionization topologies can come with a different probabilistic mapping $P_\mathrm{top}(\logNHI, \Rb \,|\, \xHI)$ between global and local labels. On the other hand, since the likelihood $L(\boldsymbol{t} \,|\, \logNHI, \Rb, t_\mathrm{Q})$ of the IGM transmission field $\boldsymbol{t}$ is reionization model independent in our local parametrization, $P_\mathrm{top}(\logNHI, \Rb \,|\, \xHI)$ is the distribution which encapsulates the \textit{entire} topology dependence of our $(\logNHI, \Rb)$ constraints.

For the sake of quoting a bare set of local, topology-independent constraints on $\logNHI$ and $\Rb$, fully agnostic to the reionization model, we initially impose a constant 
prior distribution in two-dimensional $(\logNHI, \Rb)$ parameter space which we denote as $P(\logNHI, \Rb)$ (in contrast to the physical prior $P_\mathrm{top}(\logNHI, \Rb)$ resulting from Eq.~(\ref{eq:prior})). %
This constant prior (marked by the dashed line of the $(\logNHI, \Rb)$ panel of Figure~\ref{fig:P_xHI_NHI_Rb}) is only limited by the hard physical boundaries for these two quantities as we will discuss below. 
Based on this, we infer the topology-\textit{agnostic} posterior distribution $P(\logtQ, \logNHI, \Rb \,|\, \boldsymbol{f})$ according to Bayes' theorem:
\begin{align}
\label{eq:posterior_3d_no_top}
    P&(\logtQ, \logNHI, \Rb \,|\, \boldsymbol{f}) = L(\boldsymbol{f} \,|\, \logtQ, \logNHI, \Rb) \nonumber \\
    &\times P(\logtQ, \logNHI, \Rb) / P(\boldsymbol{f}).
\end{align}
If we assume the full three-dimensional prior factorizes into $P(\logtQ, \logNHI, \Rb) = P(\logtQ) \times P(\logNHI, \Rb)$, folding in the topology dependence then simply amounts to a change of priors to the non-trivial prior $P_\mathrm{top}(\logNHI, \Rb)$ governed by the topology of interest as per Eq.~(\ref{eq:prior}):
\begin{align}
\label{eq:posterior_3d_top}
    P_\mathrm{top}&(\logtQ, \logNHI, \Rb \,|\, \boldsymbol{f}) \\
    &= \frac{P(\logtQ, \logNHI, \Rb \,|\, \boldsymbol{f})}{P(\logNHI, \Rb)} \,\times\, {P_\mathrm{top}(\logNHI, \Rb)}, \nonumber
\end{align}
where the denominator is trivial since $P(\logNHI, \Rb)$ is constant by assumption,
and it encloses the \textit{entire} physical domain, implying that the support of any non-trivial prior $P_\mathrm{top}(\logNHI, \Rb)$ is guaranteed to be a subset of the support of $P(\logNHI, \Rb)$.

We now proceed by practically determining $P_\mathrm{top}(\logNHI, \Rb)$ for the realistic semi-numerical reionization topology considered in this work. Enforcing a uniform prior on $\xHI$ is straightforward in the case at hand, as we are already provided with IGM transmission profiles at $21$ regularly spaced $\xHI$ parameter values between $0$ and $1$, well-representing a uniform prior distribution. By computing the local summary statistics $\logNHI$ and $\Rb$ for all $21 \times 2545$ sightlines (i.e., $2545$ distinct density sightlines combined with $21$ different $\xHI$ models each), 
and marginalizing over the underlying $\xHI$ values, we are therefore immediately provided with a set of samples from the prior distribution $P_\mathrm{top}(\logNHI, \Rb)$ informed by this topology.

We show in Figure~\ref{fig:P_xHI_NHI_Rb} a corner plot of the three-dimensional joint distribution $P_\mathrm{top}(\xHI, \logNHI, \Rb)$ as given by Eq.~(\ref{eq:joint}) before performing the marginalization over $\xHI$. For illustrational purposes, we depict 2d-marginals of a three-dimensional kernel density estimation (KDE) 
of the distribution rather than the actual samples, especially because of the discrete grid spacing in the $\xHI$ dimension. Note that the $(\logNHI, \Rb)$ panel of the plot can individually be understood as a contour plot 
of the prior distribution $P_\mathrm{top}(\logNHI, \Rb)$ given by Eq.~(\ref{eq:prior}) under the assumption of a flat prior on $\xHI$.

For reference, the dashed line in the $(\logNHI, \Rb)$ panel marks the two-dimensional constant prior $P(\logNHI, \Rb)$ which follows from our synthetic toy profiles used for constructing the IGM transmission likelihood. We immediately note that it covers a significantly more extended range of parameter space than the realistic ionization topology. This is because our toy prescription can produce profiles at \textit{any} parameter combination that is not excluded physically. This particularly also includes sightlines that simultaneously show low HI column densities and short neutral patch distances, i.e. the bottom left region of the $(\logNHI, \Rb)$ panel. To keep the column density low, such sightlines cannot contain many neutral patches in addition to the first one that is closest to the source demarcating the value of $\Rb$
---a configuration that is rarely found in realistic reionization topologies due to the inside-out nature of reionization. On the other hand, high HI column densities with large neutral patch distances as would be seen in the top right corner of the panel
are not found in \textit{any} topology whatsoever. This is a hard physical constraint as the maximum HI column density, set by the number of sightlines which are \textit{completely neutral} starting at $\Rb$ (and, by definition, ionized at $r < \Rb$), clearly decreases as a function of $\Rb$. 
Note that even if all material further along the line of sight 
from $\Rb$ is neutral, the density fluctuations limit the maximum value of $\logNHI$ that can be achieved. 
These maximum column density values are thus exclusively determined by the distribution of density fluctuations in the IGM.

In the same manner, we would expect the minimum column density to increase monotonically with decreasing $\Rb$, as these minimum values are set by those sightlines which are completely ionized except at a single, small neutral patch located at $\Rb$ whose contribution to $\logNHI$ increases according to the Lorentzian weighting with decreasing $\Rb$. The dashed line on the left-hand side of Figure~\ref{fig:P_xHI_NHI_Rb} shows that this is indeed the case for all neutral patch distances $150\,\mathrm{cMpc} \geq \Rb \geq 4\,\mathrm{cMpc}$. 
Note that the sharp edges we see when following the dashed line from $\Rb = 150\,\mathrm{cMpc}$ down to $\Rb = 4\,\mathrm{cMpc}$ 
result from the discrete $(\logNHI, \Rb)$ parameter grid whose spacing is chosen relatively coarsely in this region of parameter space. 

However, this behavior turns around at $\Rb < 4\,\mathrm{cMpc}$, 
where the minimum column density reduces back to $\NHI \simeq 10^{18}\,\mathrm{cm}^{-2}$ (manifesting in the kink seen in the bottom left of the $(\logNHI, \Rb)$ panel in Figure~\ref{fig:P_xHI_NHI_Rb}). 
This discontinuity arises from the fact that our lower integration limit $r_\mathrm{min}$ for the HI column density (see Eqs.~(\ref{eq:Lorentzian_avg}) and (\ref{eq:NHI_DW})) is fixed to specifically this distance. 
As a result, any neutral material 
closer to the quasar than 
$r_\mathrm{min}$ will not contribute to $\logNHI$.
To further understand the seemingly peculiar shape, now imagine two different scenarios: first, a sightline with $\Rb < r_\mathrm{min}$ which contains a single neutral patch at the location $\Rb$ closer to the quasar than $r_\mathrm{min}$ but which is ionized from $r_\mathrm{min}$ onward. Second, a sightline with $\Rb \geq r_\mathrm{min}$  which only contains a neutral patch at the location $\Rb$ more distant from the quasar than $r_\mathrm{min}$. Then the neutral patch in the second scenario with $\Rb \geq r_\mathrm{min}$ certainly contributes to $\logNHI$ while that of the first sightline does not since it is located outside of the integration range of $\logNHI$. Despite its small $\Rb$ value, the sightline in the first scenario therefore has a smaller HI column density than the sightline in the second,  whose $\Rb$ value is higher but whose neutral patch inevitably contributes to $\logNHI$. This behavior generalizes as all sightlines with $\Rb \geq r_\mathrm{min}$ by construction have at least one neutral patch contributing to $\logNHI$, and as a result, the minimum column density of sightlines with $\Rb < r_\mathrm{min}$ is \textit{smaller} than that of sightlines with $\Rb \geq r_\mathrm{min}$, giving rise to the kink we observe at $\Rb = 4\,\mathrm{cMpc}$ in Figure~\ref{fig:P_xHI_NHI_Rb}.

Note further that the realizations of the realistic topology (black dots in Figure~\ref{fig:P_xHI_NHI_Rb}) seemingly extend beyond the hard physical boundary enclosed by the dashed line. A number of effects conspire to cause this behavior. First, and most importantly, the dashed line respects the additional requirement of having available at least $800$ toy profiles to estimate smooth covariance matrices for the IGM transmission likelihood at any given point in $(\logNHI, \Rb)$ parameter space, excluding certain configurations that are rare but not entirely impossible to achieve physically. Second, as already pointed out above, the dashed line is a linear interpolation across our simulation grid which we chose rather coarsely towards lower $\logNHI$ and higher $\Rb$ values where the transmission profiles cease to vary with these parameters. Third, this behavior looks even more drastic for the probability contours shown in this figure where smoothing effects make the distribution appear wider than it in reality is.

Aside from the limited support in the $(\logNHI, \Rb)$ plane, the shape of the realistic prior $P_\mathrm{top}(\logNHI, \Rb)$ differs significantly from a uniform distribution. We observe a bimodal distribution with a more pronounced peak towards higher HI column densities of $\NHI \gtrsim 10^{20}\,\mathrm{cm}^{-2}$ and smaller neutral patch distances $\Rb \lesssim 20\,\mathrm{cMpc}$, and a lower peak at low column densities $\NHI \lesssim 10^{18.5}\,\mathrm{cm}^{-2}$ with large distances $\Rb \simeq 143\,\mathrm{cMpc}$ to the first neutral patch. 
The axis of degeneracy between the peaks naturally arises due to the fact that by definition, sightlines with a more nearby neutral patch can accommodate higher HI column densities. However, the fact that this is not a perfect one-to-one relation implies that our second summary statistic $\Rb$ captures additional information not contained in $\logNHI$.

Comparing to Figure~\ref{fig:P_NHI_Rb_given_xHI} where we show the \textit{conditional} distribution $P_\mathrm{top}(\logNHI, \Rb | \xHI)$ of the two local summaries $\logNHI$ and $\Rb$ given six different values of the global IGM neutral fraction $\xHI$, we immediately see that the former peak is due to sightlines originating from more neutral topologies, while the latter one corresponds to the most ionized ones. An axis of degeneracy connecting the two peaks is also apparent in these panels; however, there is 
significant scatter about this axis due to the fact that the distribution of neutral patches along a given line of sight can vary significantly even at fixed global IGM neutral fraction $\xHI$.
It is this amount of scatter which we exclude from the primary inference task by adopting our local parameterization rather than the global one.

Note that the high-$\Rb$ peak is artificial in the sense that our simulated skewers have a finite length of $100\,\mathrm{cMpc}/h$. Strictly speaking, $\Rb$ is ill-defined for all skewers that do not contain \textit{any} neutral patch along their entire range. In such cases, we set $\Rb$ to the maximum distance of $100\,\mathrm{cMpc}/h \simeq 143\,\mathrm{cMpc}$, noting that this value really only constitutes a lower limit for the actual distance to the first neutral patch.\footnote{Technically speaking, we are thus inferring the parameter $\Rb \equiv \min(\Rb^\mathrm{phys}, 143\,\mathrm{cMpc})$, where $\Rb^\mathrm{phys}$ is the true physical distance to the first neutral patch.} However, due to the Lorentzian decline of the Lyman-$\alpha$ cross section $\sigma_\alpha$, such distant neutral patches hardly leave any imprint on the observed IGM transmission profile anyways, and therefore this artificial bound will not bias our constraints.

\subsubsection{Converting local measurements to a global $\xHI$ constraint}
\label{sec:conversion}

Due to their local nature, our two summary statistics encapsulate information not only about the global timing of reionization, but also the local ionization topology itself. The former bit is conventionally quoted in terms of the global volume-averaged IGM neutral fraction $\xHI$. Here we elaborate on how such a global constraint can easily be obtained from our local summary statistics by again invoking the stochastic
relation $P_\mathrm{top}(\logNHI, \Rb \,|\, \xHI)$ between global and local parameters. Apart from setting the prior, the topology dependence therefore naturally comes in a second time when converting the local $(\logNHI, \Rb)$ constraints to a global constraint on $\xHI$.

Specifically, let us assume that, given an observed quasar spectrum $\boldsymbol{f}$, we inferred the two local summaries $(\logNHI, \Rb)$ as well as the lifetime of the quasar $\logtQ$, i.e., we obtained samples from the local, topology-agnostic posterior distribution $P(\logtQ, \logNHI, \Rb \,|\, \boldsymbol{f})$ via Eq.~(\ref{eq:posterior_3d_no_top}), 
but we are interested in the full topology-informed posterior distribution $P_\mathrm{top}(\xHI, \logtQ, \logNHI, \Rb \,|\, \boldsymbol{f})$ of both global and local parameters. By decomposing the latter via Bayes' theorem, we can see that
\begin{align}
\label{eq:posterior_4d}
    P_\mathrm{top}&(\xHI, \logtQ, \logNHI, \Rb \,|\, \boldsymbol{f}) \\
    &= L_\mathrm{top}(\boldsymbol{f} \,|\, \xHI, \logtQ, \logNHI, \Rb) \nonumber \\
    &\;\;\;\; \times P_\mathrm{top}(\xHI, \logtQ, \logNHI, \Rb) / P(\boldsymbol{f}). \nonumber
\end{align}
We can simplify this expression by harnessing the topology-independence of the IGM transmission likelihood in our local parameterization, which implies $L_\mathrm{top}(\boldsymbol{f} \,|\, \xHI, \logtQ, \logNHI, \Rb) = L(\boldsymbol{f} \,|\, \logtQ, \logNHI, \Rb)$ as the shape of the transmission profiles is not affected by what topology of what global neutral fraction $\xHI$ they originate from, provided that the values of the local summary statistics $\logNHI$ and $\Rb$ are fixed. By further assuming an independent lifetime prior $P(\logtQ)$ such that $P_\mathrm{top}(\xHI, \logtQ, \logNHI, \Rb) = P(\logtQ) \times P_\mathrm{top}(\xHI, \logNHI, \Rb)$, and decomposing the remaining joint distribution $P_\mathrm{top}(\xHI, \logNHI, \Rb)$ as per Eq.~(\ref{eq:joint}), 
we can straightforwardly combine Eqs.~(\ref{eq:posterior_3d_no_top}) and (\ref{eq:posterior_4d})
to replace the likelihood $L(\boldsymbol{f} \,|\, \logtQ, \logNHI, \Rb)$ with the topology-agnostic posterior $P(\logtQ, \logNHI, \Rb | \boldsymbol{f})$ and arrive at
\begin{align}
\label{eq:posterior_final}
    P_\mathrm{top}&(\xHI, \logtQ, \logNHI, \Rb | \boldsymbol{f}) \!=\! \frac{P(\logtQ, \logNHI, \Rb | \boldsymbol{f})}{P(\logNHI, \Rb)} \nonumber \\
    &\times P_\mathrm{top}(\logNHI, \Rb \,|\, \xHI)\,\times\, P(\xHI).
\end{align}
Note that the denominator is trivial since it only consists of the constant topology-agnostic prior $P(\logNHI, \Rb)$ that covers the entire physical domain (dashed line in Figures~\ref{fig:P_xHI_NHI_Rb} and \ref{fig:P_NHI_Rb_given_xHI}). In essence, folding in the topology dependence to obtain the full four-dimensional topology-informed posterior distribution $P_\mathrm{top}(\xHI, \logtQ, \logNHI, \Rb \,|\, \boldsymbol{f})$ is thus simply a matter of multiplying together the local, topology-agnostic posterior $P(\logtQ, \logNHI, \Rb \,|\, \boldsymbol{f})$ that we originally inferred, and the conditional probability distribution $P_\mathrm{top}(\logNHI, \Rb \,|\, \xHI)$ which encapsulates \textit{all} information about the assumed ionization topology, along with the prior $P(\xHI)$. In Eq.~(\ref{eq:posterior_final}), we are therefore 1) imposing a non-trivial, physically motivated prior distribution on our local parameters, and 2) obtaining a constraint on the global timing of reionization.

In practice, we have to keep in mind that we are only provided with \textit{samples} from the two aforementioned distributions. In order to perform the multiplication in Eq.~(\ref{eq:posterior_final}), we introduce a parameter grid to evaluate the distributions on. We choose the $(\xHI, \logtQ)$ grid spacing in accordance with our parameter grid for the IGM transmission profiles discussed in Section~\ref{sec:sims}, and likewise for $(\logNHI, \Rb)$ based on the irregular parameter grid on which our toy transmission profiles used to determine the IGM transmission likelihood are simulated.\footnote{Adopting this irregular binning ensures that we make use of the enhanced sensitivity for profiles with strong damping wings, i.e. at high $\logNHI$ and low $\Rb$, where we chose a finer grid spacing. We formally investigate the precision of our measurements as a function of astrophysical parameter space in Section~\ref{sec:precision}.} We then represent the two distributions through histograms with bins defined by these grids. As soon as $P_\mathrm{top}(\xHI, \logtQ, \logNHI, \Rb \,|\, \boldsymbol{f})$ is determined according to Eq.~(\ref{eq:posterior_final}), we can arbitrarily marginalize the distribution via numerical integration over our parameter grid in order to obtain, e.g., the conventionally quoted posterior $P_\mathrm{top}(\xHI, \logtQ \,|\, \boldsymbol{f})$ on the global IGM neutral fraction $\xHI$ and quasar lifetime $\logtQ$, or the associated one-dimensional marginals $P_\mathrm{top}(\xHI \,|\, \boldsymbol{f})$ or $P_\mathrm{top}(\logtQ \,|\, \boldsymbol{f})$.

\section{Methods: inference pipeline}
\label{sec:methods}

We described in the previous section a set of summary statistics that encodes the local information encapsulated in the IGM damping wing imprint in a topology-independent fashion. We now proceed by introducing an inference framework that can be used to constrain these parameters based on observed high-redshift quasar spectra. The formalism is based on the one established in \citet{hennawi2025} and \citet{kist2025a} as a fully Bayesian framework to infer the global IGM neutral fraction $\xHI$ and the lifetime $\logtQ$ of a quasar based on its observed spectrum $\boldsymbol{f}$. We start this section with a short summary providing an overview over all major modeling components, focusing especially on the adaptations made to constrain our local summary statistics rather than directly inferring the global IGM neutral fraction $\xHI$. For specific details, we point the reader to \citet{hennawi2025}.

The damping wing signature is imprinted upon the spectra of high-redshift sources (such as quasars in our case) by the foreground IGM. In order to extract the information it encapsulates about the global---or local---ionization topology, we therefore first have to disentangle this imprint from the unabsorbed, intrinsic spectrum of the source. A two-step method has been the conventional approach to address this task \citep[see e.g.][]{greig2017a, davies2018a}: first, the intrinsic continuum of the quasar is reconstructed based on its correlation with the emission lines in the unabsorbed region of the spectrum redward of the Lyman-$\alpha$ line. \textit{Subsequently}, the resulting continuum-normalized spectrum can be used to draw conclusions about the ionization state of the surrounding IGM. \citet{hennawi2025} introduced for the first time a fully Bayesian framework that \textit{jointly} performs these two tasks while operating on the entire spectral range, thus accounting for the full covariance resulting from the continuum reconstruction and the IGM transmission stochastic process.

The heart of the framework is a full generative model for high-redshift quasar spectra $\boldsymbol{f}$. Starting from a quasar continuum $\boldsymbol{s}$, based on a dataset of $\sim 45\,000$ unabsorbed, low-redshift ($1.878 < z < 3.427$) continua, we fold in the IGM absorption imprint based on simulated IGM transmission profiles $\boldsymbol{t}$ (see Section~\ref{sec:sims}), parameterized as a function of $\boldsymbol{\theta} \equiv (\logtQ, \logNHI, \Rb)$. After forward-modeling instrumental effects by convolving these profiles with the line-spread function of a hypothetical spectrograph, 
as well as adding a realistic, heteroscedastic spectral noise vector $\boldsymbol{\sigma}$, we are equipped with a realistic forward model for high-redshift quasar spectra $\boldsymbol{f}$.

To tackle the inverse problem of constraining the astrophysical parameters $\boldsymbol{\theta}$ based on an observed quasar spectrum $\boldsymbol{f}$, we construct a data-driven low-dimensional parametric PCA model for the quasar continuum based on the aforementioned set of low-redshift spectra. Together with the simulated IGM transmission profiles $\boldsymbol{t}$, we use this to determine the full likelihood $L(\boldsymbol{f} \,|\, \boldsymbol{\theta}, \boldsymbol{\eta}, \boldsymbol{\sigma})$ of the observed quasar spectrum $\boldsymbol{f}$ as a function of the low-dimensional vector $\boldsymbol{\eta}$ of (latent) PCA coefficients describing the full continuum $\boldsymbol{s}$ and the astrophysical parameter vector $\boldsymbol{\theta}$, conditioned on the observational noise vector $\boldsymbol{\sigma}$. \citet{hennawi2025} demonstrated that we can approximate this likelihood as
\begin{equation}
\label{eq:likelihood}
    L(\boldsymbol{f}|\boldsymbol{\sigma}, \boldsymbol{\theta}, \boldsymbol{\eta}) = \mathcal{N}\left(\boldsymbol{f} ;\langle\boldsymbol{t}\rangle \circ\langle\boldsymbol{s}\rangle, \boldsymbol{\Sigma}+\langle\boldsymbol{S}\rangle \boldsymbol{C}_{\boldsymbol{t}}\langle\boldsymbol{S}\rangle+\langle\boldsymbol{T}\rangle \boldsymbol{C}_{\boldsymbol{s}}\langle\boldsymbol{T}\rangle\right),
\end{equation}
where $\mathcal{N}(\boldsymbol{f}; \boldsymbol{\mu}, \boldsymbol{K})$ is the multivariate normal distribution of the random variable $\boldsymbol{f}$ with mean $\boldsymbol{\mu}$ and covariance matrix $\boldsymbol{K}$. Further, $\boldsymbol{t}\circ\boldsymbol{s}$ denotes the element-wise (Hadamard) product of the two mean vectors $\langle\boldsymbol{t}\rangle$ and $\langle\boldsymbol{s}\rangle$, and we defined the matrices $\boldsymbol{\Sigma} \equiv \mathrm{diag}(\boldsymbol{\sigma})$,  $\boldsymbol{T}\equiv\mathrm{diag}(\boldsymbol{t})$, $\boldsymbol{S}\equiv\mathrm{diag}(\boldsymbol{s})$, as well as the covariance matrices $\boldsymbol{C}_{\boldsymbol{t}}$ and $\boldsymbol{C}_{\boldsymbol{s}}$ of $\boldsymbol{t}$ and $\boldsymbol{s}$, respectively.

Note that this likelihood operates on the entire spectral range, both redward and blueward of the Lyman-$\alpha$ line, and hence covers the smooth IGM damping wing as well as the Lyman-$\alpha$ forest region with the quasar proximity zone. This allows us to \textit{jointly} infer the astrophysical parameters $\boldsymbol{\theta}$ and the continuum nuisance parameters $\boldsymbol{\eta}$.

\subsection{Continuum dimensionality reduction model}
\label{sec:continuum}

We now proceed with a short summary of our parametric model for the quasar continuum. In short, we are using an updated version of the principal component analysis (PCA) model described in \citet{hennawi2025} which additionally accounts for luminosity variations in the spectra. This allows us to construct the PCA decomposition based on a $\sim 3$ times larger dataset of low-redshift continua, and extend the spectral coverage redward of Lyman-$\alpha$ out to the Mg II line. A detailed description of this new model will be provided in \citet{hennawi2025b}.

In short, the dataset we use comprises $44,587$ low-redshift ($1.878 < z < 3.427$) spectra from the SDSS-III Baryon Oscillation Spectroscopic Survey (BOSS) and SDSS-IV Extended BOSS (eBOSS).
These spectra cover a rest-frame wavelength range of $1175 - 3000\,\text{\AA}$ and have a resolution of $R\sim2000$ with a median signal-to-noise ratio of $\mathrm{S}/\mathrm{N} > 10$ within a $5\,\text{\AA}$ region around the rest-frame wavelength $1285\,\text{\AA}$. 
For a small subset of these sources, we also obtained near-infrared spectra from the Gemini Near-Infrared Spectrograph Distant Quasar Survey \citep[GNIRS-DQS;][]{matthews2021}. The spectra were taken with the Gemini Near-Infrared Spectrograph \citep[GNIRS;][]{elias2006} at the Gemini North Observatory. These spectra cover the $\sim0.8{-}2.5\mu$m range to encompass the H$\beta$ and [\ion{O}{iii}] region. They have a resolution of $R \sim 1100$ and were obtained to reach sensitivities comparable to the SDSS spectra at $\lambda_{\rm obs}\sim5000~\text{\AA}$. All spectroscopic data from GNIRS-DQS were re-reduced with PypeIt \citep{prochaska2020}.

We compute the PCA decomposition based on $42,854$ (i.e., $\sim 96\,\%$) of the full set of continua and keep the remaining $1733$ apart as test set for estimating the reconstruction error and drawing mock continua. The PCA decomposition is computed based on the logarithmic continua, and we keep the $7$ PCA vectors accounting for the largest variance as basis vectors of our model. Note that since we perform the decomposition in log-space, the first PCA vector captures the normalization of the continua, and hence the luminosity dependence of the spectral shape is intrinsically accounted for. Note further that we opted for a weighted PCA that allows us to also include continua with missing pixels. For convenience, however, we chose our test set such that it exclusively contains continua with full spectral coverage. This provides us with a dimensionality-reduced description $\boldsymbol{s}_\mathrm{DR}(\boldsymbol{\eta})$ of the true continuum $\boldsymbol{s}$ with the model parameters corresponding to the seven PCA coefficients $\boldsymbol{\eta}$.\footnote{As demonstrated in \citet{kist2025a}, a five-dimensional PCA model plus normalization factor is sufficiently flexible to represent the full quasar continuum $\boldsymbol{s}$ for the purpose of astrophysical parameter inference based on IGM damping wings. As we perform the PCA in log-space without the need for an additional normalization factor, and due to the extended red-side spectral coverage of $3000\,\text{\AA}$ \citep[as compared to $2000\,\text{\AA}$ in the analysis by][]{kist2025a}, we increase the dimensionality of our PCA model to seven.}

\citet{hennawi2025} showed that the associated relative continuum reconstruction error $\boldsymbol{\delta} = (\boldsymbol{s}-\boldsymbol{s}_\mathrm{DR})/\boldsymbol{s}$ is well-approximated by a Gaussian distribution, and, as a result,
\begin{equation}
\label{eq:P_s_eta}
    P(\boldsymbol{s}|\boldsymbol{\eta}) = \mathcal{N}(\boldsymbol{s}; \langle\boldsymbol{s}(\boldsymbol{\eta})\rangle, \boldsymbol{C}_{\boldsymbol{s}}(\boldsymbol{\eta})).
\end{equation}
We estimate the mean $\langle\boldsymbol{s}(\boldsymbol{\eta})\rangle \equiv \boldsymbol{s}_\mathrm{DR}(\boldsymbol{\eta})\circ(1+\langle\boldsymbol{\delta}\rangle)$ and the covariance $\boldsymbol{C}_{\boldsymbol{s}}(\boldsymbol{\eta}) \equiv \mathrm{diag}(\boldsymbol{s}_\mathrm{DR}(\boldsymbol{\eta}))\;\boldsymbol{C}_{\boldsymbol{\delta}}\;\mathrm{diag}(\boldsymbol{s}_\mathrm{DR}(\boldsymbol{\eta}))$ of this distribution based on the test set of $1733$ 
continua, where $\langle\boldsymbol{\delta}\rangle$ and $\boldsymbol{C}_{\boldsymbol{\delta}}$ are the mean and covariance of the continuum reconstruction error $\boldsymbol{\delta}$.

\subsection{IGM transmission likelihood}
\label{sec:transm_likelihood}

All patches of neutral hydrogen in the surrounding IGM imprint an absorption signature on the continuum of the quasar. This is represented by IGM transmission profiles $\boldsymbol{t}$ which we obtain from numerical simulations (see Section~\ref{sec:sims}), parameterized by a set of astrophysical parameters $\boldsymbol{\theta}$. In the conventional approach, one adopts $\boldsymbol{\theta} \equiv (\xHI, \logtQ)$ which we henceforth refer to as the \textit{global} parameterization, whereas we here propose to consider the \textit{local} parameterization $\boldsymbol{\theta} \equiv (\logtQ, \logNHI, \Rb)$.

In either case, to go forward, we follow \citet{hennawi2025} and approximate the distribution of $\boldsymbol{t}$ given the astrophysical parameters $\boldsymbol{\theta}$ as a multivariate Gaussian distribution:
\begin{equation}
\label{eq:P_t_theta}
    P(\boldsymbol{t}|\boldsymbol{\theta}) = \mathcal{N}(\boldsymbol{t}; \langle\boldsymbol{t}(\boldsymbol{\theta})\rangle, \boldsymbol{C}_{\boldsymbol{t}}(\boldsymbol{\theta})).
\end{equation}
It is a well-known fact that the assumption of Gaussianity is in fact not fully justified when using $\xHI$ as label \citep{lee2015, davies2018a}, and neither in the case where $\xHI$ gets replaced with our local summary statistics as we find in this analysis. The approximation, however, is vital to our formalism as it allows us to derive a closed-form analytical expression for the likelihood of the full quasar spectrum. To make sure that this does not result in an overestimation of our constraining power, we apply a principled procedure to broaden the posteriors by the degree required to ensure statistically faithful constraints \citep[see also][]{hennawi2025}. In future, we aim to address these issues while retaining the full constraining power by learning the true non-Gaussian shape of $P(\boldsymbol{t}|\boldsymbol{\theta})$ with the help of simulation-based inference.

As far as this work is concerned, we simply estimate the mean transmission $\langle\boldsymbol{t}(\boldsymbol{\theta})\rangle$ and the covariance $\boldsymbol{C}_{\boldsymbol{t}}(\boldsymbol{\theta})$ based on our simulated IGM transmission profiles. Note that sampling from the posterior distribution via Hamiltonian Monte Carlo (HMC) becomes challenging in cases where the posterior distribution is not a smooth function of the parameters $\boldsymbol{\theta}$ as this can prevent sufficient mixing of the HMC chains. The most critical part of our framework are the covariance matrices $\boldsymbol{C}_{\boldsymbol{t}}(\boldsymbol{\theta})$ which are prone to becoming noisy if estimated based on too low a number of sightlines. This potential pitfall is addressed by our synthetic procedure for generating $x_\mathrm{HI}$ sightlines that vary smoothly as a function of $\logNHI$ and $\Rb$, as described in Section~\ref{sec:sims} \citep[see also][]{kist2025b}. This allows us to measure means and covariances based on a large number of sightlines at any desired location in $(\logNHI, \Rb)$ parameter space with continuity with respect to those parameters intrinsically built in. An additional virtue of this procedure is that it allows us to cover the \textit{entire} physically accessible domain in $(\logNHI, \Rb)$ parameter space with a sufficient number of simulated sightlines for estimating covariance matrices.

Note also that before computing $\langle\boldsymbol{t}(\boldsymbol{\theta})\rangle$ and  $\boldsymbol{C}_{\boldsymbol{t}}(\boldsymbol{\theta})$, we forward-model instrumental effects for each individual transmission profile by convolving it with a Gaussian line-spread function (LSF) of width $\mathrm{FWHM} = 100\,\mathrm{km}/\mathrm{s}$ and rebinning it onto a coarse velocity grid with a pixel spacing of $500\,\mathrm{km/\mathrm{s}}$, covering a rest-frame wavelength range of $1175 - 3000\,\text{\AA}$. Even though strictly speaking this should not be done before multiplying in the quasar continuum, we perform these operations in \textit{pre-processing} on the bare IGM transmission profiles, as doing this downstream would make the inference computationally infeasible without recognizable precision gains.

\subsection{Inference procedure}

With a full framework at hand for the likelihood $L(\boldsymbol{f}|\boldsymbol{\sigma}, \boldsymbol{\theta}, \boldsymbol{\eta})$ of an observed quasar spectrum $\boldsymbol{f}$ given astrophysical parameters $\boldsymbol{\theta}$ and continuum nuisance parameters $\boldsymbol{\eta}$, we are now in the position to sample from the associated posterior distribution once we specified our priors. We adopt a uniform prior on the logarithmic quasar lifetime $P(\logtQ) = \mathrm{Unif}(3, 8)$, covering a wide range of physically plausible values \citep{khrykin2021}. In the context of our local parameter framework, we impose a two-dimensional constant prior
$P(\logNHI, \Rb)$ on our local summary statistics, enclosed by the (non-trivial) boundaries set by the distribution of IGM density fluctuations as discussed in Section~\ref{sec:prior}. For reference, we also perform the inference in the conventional, global $(\xHI, \logtQ)$ framework, in which case we impose an uninformative, uniform prior on the IGM neutral fraction $P(\xHI) = \mathrm{Unif}(0, 1)$ in addition to the aforementioned log-uniform lifetime prior. In practice, to aid the sampling procedure, we apply sigmoid transformations to all bounded parameters to make them fully unbounded. To account for the non-trivial two-dimensional prior boundary of $P(\logNHI, \Rb)$, we first transform $\Rb$, and subsequently $\logNHI$, conditioned on $\Rb$. This is required since the minimum and maximum HI column densities $\logNHI$ change as a function of $\Rb$. The PCA coefficients $\eta_i$ parameterizing the shape of the quasar continuum are already allowed to vary over the entire real axis $P(\eta_i) = \mathrm{Unif}(-\infty, \infty)$, so here we only remove their mean and rescale them by their standard deviation.

We then deploy Hamiltonian Monte-Carlo (HMC) to sample from the resulting posterior distribution, using the HMC implementation with a No U-Turn Sampler (NUTS) from the \texttt{NumPyro} probabilisitic programming library based on the machine learning and autograd framework \texttt{JAX} \citep{bradbury2018, bingham2018, phan2019}. For a given spectrum, we run four
HMC chains with $1000$ warm-up steps each, and an additional $1000$ steps for sampling. %
Furthermore, we perform sigma-clipping on the spectrum before performing the inference itself in order to make sure our results do not get biased by a small number of outlier pixels. To that end, we start by iteratively fitting for the maximum-likelihood model with the help of an Adam optimizer with a learning rate of $0.5$ and $1000$ optimization steps. We compute the $\chi^2$ statistic between the data and this best-fit model, masking up to five $> 4\,\sigma$ outlier pixels on the red side of the spectrum ($\lambda_\mathrm{rest} > 1260\,\text{\AA}$) per iteration. 
We repeat the same optimizing and clipping procedure on the remaining unmasked part of the spectrum for up to $5$ iterations before we start sampling from the posterior distribution via HMC.

\subsection{Generating mock spectra}
\label{sec:mocks}

To verify the performance of our pipeline, we apply it to mock spectra generated in the following way. We start by drawing an unabsorbed continuum $\boldsymbol{s}$ from the test set of $1733$ low-redshift 
spectra used in Section~\ref{sec:continuum} to estimate the continuum reconstruction error. Given a desired astrophysical parameter vector $\boldsymbol{\theta}$, we further draw an IGM transmission profile $\boldsymbol{t}$ from the simulated profiles described in Section~\ref{sec:sims}, with instrumental effects folded in as discussed in Section~\ref{sec:transm_likelihood} via convolution with a Gaussian LSF of width $\mathrm{FWHM} = 100\,\mathrm{km}/\mathrm{s}$ and rebinning to a $500\,\mathrm{km/\mathrm{s}}$ velocity grid. We interpolate the continuum $\boldsymbol{s}$ onto the same wavelength grid (extending from $1175\,\text{\AA}$ to $3000\,\text{\AA}$), and multiply it together with the IGM transmission profile $\boldsymbol{t}$. Finally, we generate realistic heteroscedastic noise realizations including telluric absorption as well as contributions from object photons, sky background and detector read noise with help of the \texttt{SkyCalc\_ipy} package \citep{leschinski2021}; see \citet{hennawi2025} for details. 
The exposure time of the hypothetical instrument is adjusted such that the median signal-to-noise ratio per $100\,\mathrm{km}/\mathrm{s}$ velocity interval is $\mathrm{S}/\mathrm{N} = 10$ within a $5\,\text{\AA}$ interval centered at $1285\,\text{\AA}$ in the rest-frame.
Two different ensembles of such mock spectra are considered in this work:
\begin{enumerate}
    \item A set of $10 \times 10$ mock spectra on a parameter grid spanned by the global IGM neutral fraction $\xHI$ and the quasar lifetime $\logtQ$ with values of $\xHI = 0.05, 0.15, ..., 0.95$ and $\logtQ = 3.25, 3.75, ..., 7.75$. The underlying IGM transmission profiles originate from the realistic semi-numerical reionization topology described in Section~\ref{sec:sims}. For each sightline, we can also uniquely determine the values of our two local pre-quasar summary statistics $\logNHI$ and $\Rb$. By construction of this sample, the distribution of these statistics follows the topology-informed prior $P_\mathrm{top}(\logNHI, \Rb)$. We run both the global and the local version of our inference pipeline on these spectra, inferring either the global parameters $(\xHI, \logtQ)$ or the local ones $(\logtQ, \logNHI, \Rb)$. We use this mock ensemble to compare the fits to these spectra in the context of the two different parameterizations, to perform coverage tests, and to compute the corrections required to pass them, as will be discussed in the subsequent section.
    \item A set of $360$ mock spectra on a $3 \times 4 \times 4$ parameter grid in the local $(\logtQ, \logNHI, \Rb)$ parameterization with values of $\logtQ = 4, 6, 8$, $\logNHI = 20.1, 20.3, 20.5, 20.7$ and $\Rb = 2, 6, 10, 14\,\mathrm{cMpc}$. At each location in parameter space, we simulate $10$ distinct mocks, and the keep the continuum and noise realization of these $10$ mock spectra fixed when varying $\logtQ$, $\logNHI$ and $\Rb$ to isolate the impact of these three parameters from those additional sources of stochasticity. 
    Note that realizations of IGM transmission profiles only exist at $12$ out of the $16$ locations in $(\logNHI, \Rb)$ parameter space since some parameter combinations are physically excluded by the distribution of density fluctuations in the IGM (see e.g. Figure~\ref{fig:P_xHI_NHI_Rb}), and hence we end up with $360$ rather than $480$ spectra. The underlying IGM transmission profiles originate from synthetic neutral fraction sightlines generated according to our analytical prescription described in Section~\ref{sec:sims}. As such, they 
    do not have an associated 
    global IGM neutral fraction value $\xHI$. We therefore only run the local version of our inference pipeline on these spectra to determine the precision with which we can infer $\logtQ$, $\logNHI$ and $\Rb$.
\end{enumerate}

\subsection{Coverage tests}
\label{sec:inf_test_theory}

A reliable inference framework guarantees that the inferred posterior distribution is statistically faithful. For example, when inferring parameters from $100$ different mock spectra, and considering the $68\,\%$ highest-density region (HDR) of each inferred posterior, 
we statistically expect the true value to be contained in this $68\,\%$ HDR for $68$ of these $100$ mocks, and likewise for any other credibility level $\alpha$. More generally speaking, given a set of $N_\mathrm{mock}$ mock inferences, we can determine the \textit{expected coverage probability} $C_\alpha$ corresponding to the \textit{credibility level} $\alpha$ as the number of mocks whose true parameter value is indeed contained in the $\alpha$-th HDR of the posterior distribution, divided by the total number of mocks in the sample.

We can perform a \textit{coverage test} by explicitly computing $C_\alpha$ for values of $\alpha$ in the entire range $\alpha \in [0, 1]$. This test is passed if we find $C_\alpha = \alpha$ for all $\alpha \in [0, 1]$, implying that the inferred posteriors contain the truth exactly as many times as they are statistically expected to. On the other hand, finding $C_\alpha > \alpha$ indicates that the inferred posteriors are \textit{underconfident}, i.e., less constraining than they could be.
\textit{Overconfident}, or too narrow posteriors where $C_\alpha < \alpha$ pose an even larger problem as they can lead us to draw conclusions that are not actually backed by the data. This is caused by flaws in the inference pipeline, such as bugs, inappropriate priors, or, as in the case at hand, an approximate likelihood prescription.

Eliminating such imperfections is essential before quoting any physical parameter constraints. While it would exceed the scope of this work to restructure our pipeline with the help of simulation-based inference to avoid the approximate analytical likelihood expression in Eq.~(\ref{eq:likelihood}), \citet{hennawi2025} introduced a principled way of retrospectively broadening the inferred posteriors in such a way that they are guaranteed to pass a coverage test. In essence, the idea is to relabel the $\alpha$-th HDR of a given posterior distribution as the $C_\alpha$-th HDR according to the mapping $\alpha \mapsto C_\alpha$ determined in the coverage test. Doing so for all credibility levels $\alpha$ by construction ensures perfect coverage for the new coverage-corrected posterior distribution. While \citet{hennawi2025} outlined the practical procedure for the case where we are provided with MCMC samples from the posterior distribution, we here elaborate on an analogous strategy applicable in the case where the posterior is evaluated on a parameter grid such as the one we are faced with when converting our local parameter constraints to global ones.

Specifically, suppose that the full parameter space spanned by $\boldsymbol{\theta}$ is divided into grid cells $i$ with central value $\boldsymbol{\theta}_i$ and volume $V_i$. Then the probability mass contained in the $i$-th pixel is
\begin{equation}
    P_i \equiv \int_{V_i} \mathrm{d}\boldsymbol{\theta}\,P(\boldsymbol{\theta}) \simeq V_i \cdot P(\boldsymbol{\theta}_i),
\end{equation}
where for clarity we suppressed all other variables in the argument of the posterior distribution $P(\boldsymbol{\theta})$. To determine the highest-density regions, we can now order all grid cells according to the probability mass they contain, resulting in a permutation $\pi(i)$ of the original grid cells $i$, with $\pi^{-1}(j)$ mapping back the sorted grid cells $j$ to their original index. We then define the $\alpha_N$-th HDR as the region enclosed by the $N$ highest-probability cells, i.e., %
\begin{equation}
    \sum_{j=1}^{N} P_{\pi^{-1}(j)} = \alpha_N.
\end{equation}
Note that this can be seen as the cumulative distribution function of the reordered cells. This also underlines that summing over all $N_\mathrm{cell}$ cells, we obtain $\sum_{j=1}^{N_\mathrm{cell}} P_{\pi^{-1}(j)} = \alpha_{N_\mathrm{cell}} = 1$.

To perform a coverage test, we can now easily determine for each mock the index $N_0$ as the smallest 
index $N$ for which the true parameter is still contained in its $\alpha_N$-th HDR.
The coverage probability corresponding to the credibility level $\alpha$ is then simply given by the relative number of objects whose value of $\alpha_{N_0}$ (i.e., the size of the smallest HDR that still contains the truth) is less or equal to $\alpha$. In other words, by rank-ordering the $N_\mathrm{mock}$ values of $\alpha_{N_0}$, and dividing their rank by the total number of mocks $N_\mathrm{mock}$, we are immediately provided with the corresponding coverage probabilities $C_{\alpha_{N_0}}$. By interpolating $C_\alpha$ among intermediate $\alpha$ values, 
we can obtain a smooth approximation of the coverage curve across the entire range $\alpha \in [0, 1]$.

Once this curve is determined, we can also make use of it to correct for any potential imperfections in the coverage behavior. This can be straightforwardly accomplished in a backward approach where we again loop over all grid cells $N$ ordered by their probability mass $P_{\pi^{-1}(N)}$, and instead assign them the \textit{correct} mass $P^\mathrm{rew}_{\pi^{-1}(N)}$ dictated by their expected coverage probability $C_{\alpha_N}$. Specifically, when considering the $N$-th cell, we have to set
\begin{equation}
    P^\mathrm{rew}_{\pi^{-1}(N)} \equiv C_{\alpha_N} - \sum_{j=1}^{N-1} P^\mathrm{rew}_{\pi^{-1}(j)}.
\end{equation}
Starting at the cell $N=1$ containing the highest probability mass, and subsequently moving to lower-probability cells $j$, this uniquely defines the values of the coverage-corrected posterior distribution $P^\mathrm{rew}(\boldsymbol{\theta})$ at each grid cell.

\section{Results: global vs local parameter inference}
\label{sec:inference_results}

Equipped with our new local parameterization of quasar IGM damping wings (Section~\ref{sec:theory}), as well as an inference framework to constrain them (Section~\ref{sec:methods}), we now proceed by testing the statistical fidelity of this framework on large ensembles of mock spectra and quantifying the precision with which we can constrain both the local and global astrophysical parameters. %
Importantly, we also demonstrate how we can relate our local constraints back to the global IGM neutral fraction $\xHI$, obtaining the full four-dimensional, topology-informed posterior $P(\xHI, \logtQ, \logNHI, \Rb | \boldsymbol{f})$ on both global and local parameters.

\subsection{Inference for a single mock}

\begin{figure*}
	\includegraphics[width=\textwidth]{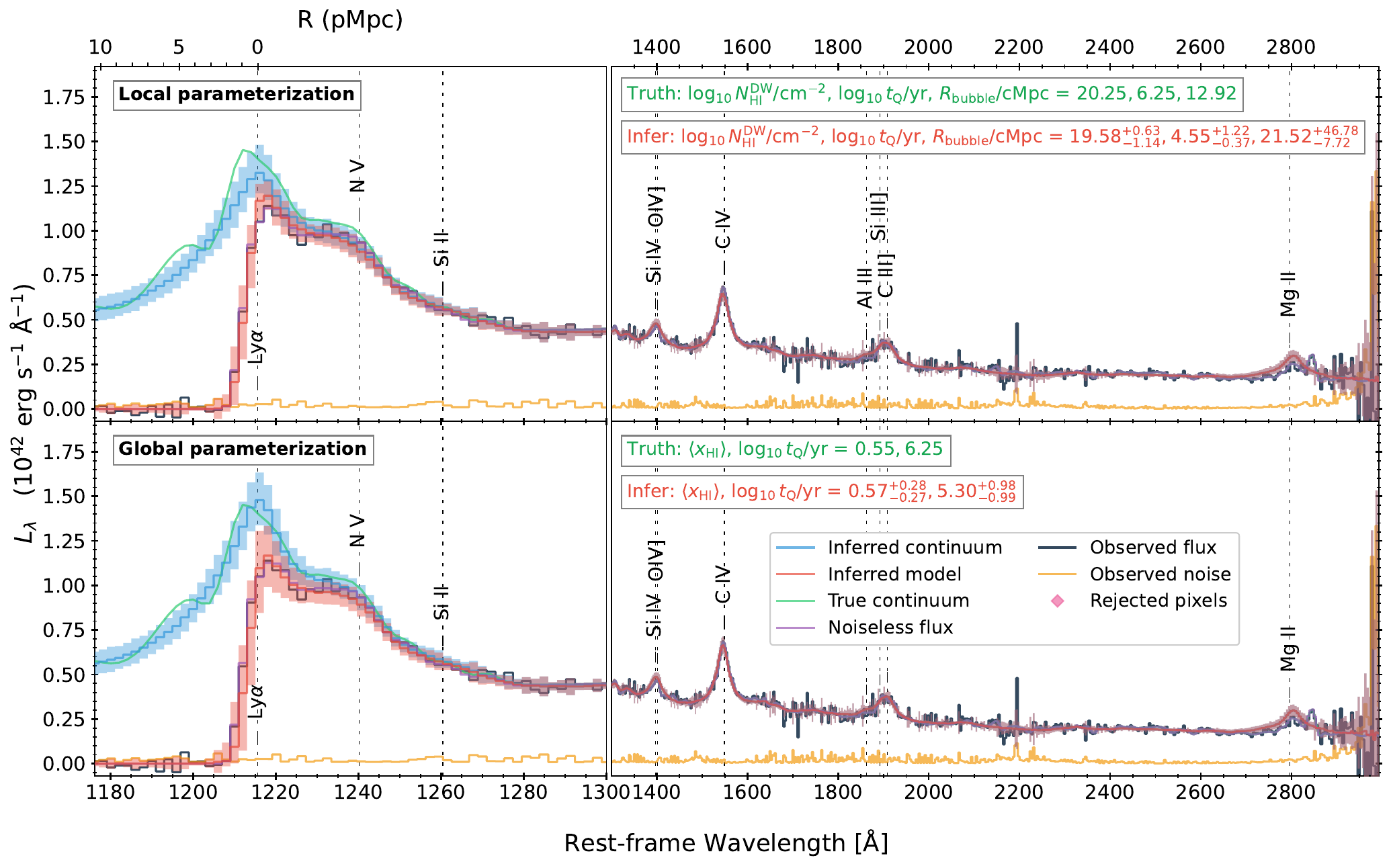}
    \caption{Inferred model for a mock spectrum of a quasar that has been shining for $\tQ = 10^{6.25}\,\mathrm{yr}$ in a globally $\xHI = 0.55$ neutral IGM, fitted with the local IGM damping wing parameterization (upper row) and the global one (lower row). The underlying pre-quasar sightline has a HI column density of $\NHI = 10^{20.25}\,\mathrm{cm}^{-2}$ and a distance to the first neutral patch of $\Rb = 12.9\,\mathrm{cMpc}$. The mock spectrum of the quasar is depicted in black and consists of the true continuum (green) including IGM absorption (purple) and spectral noise (yellow). The inferred model spectrum is depicted in red, with the unabsorbed inferred continuum shown in blue. Solid lines represent the median inferred models, shaded regions the $16\,\%$ and the $84\,\%$ percentile variations reflecting parameter uncertainty, continuum reconstruction errors, as well as spectral noise.}
    \label{fig:mock_spec}
\end{figure*}

\begin{figure*}
	\includegraphics[width=\textwidth]{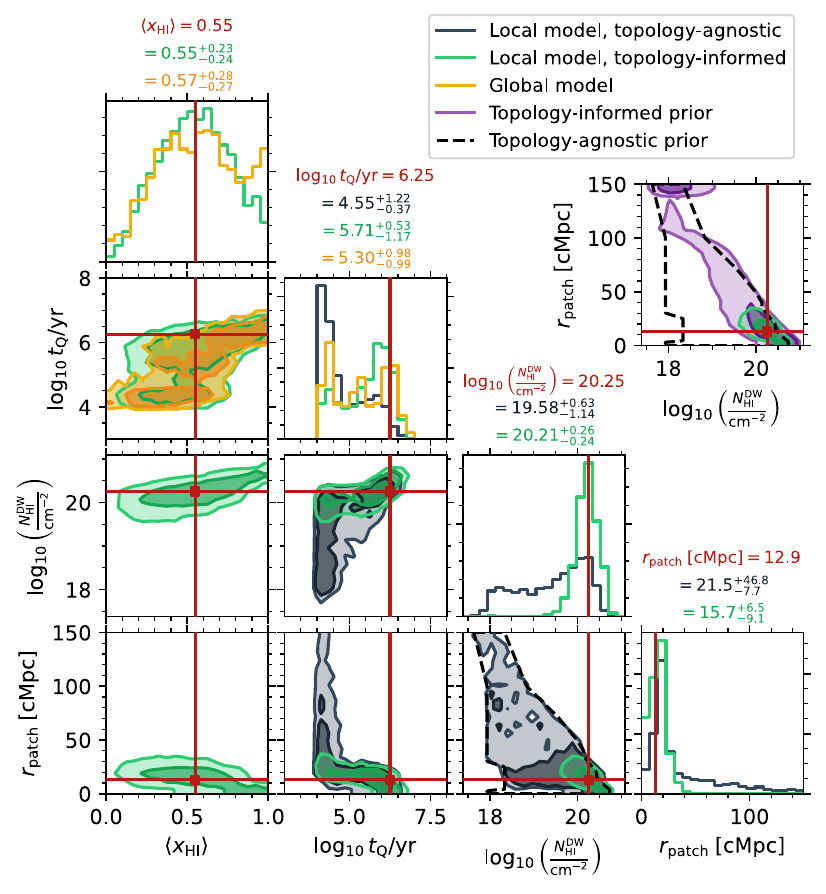}
    \caption{Posterior distributions inferred from the mock spectrum depicted in Figure~\ref{fig:mock_spec} in the local IGM damping wing parameterization (black) and the global one (yellow). In both cases, the distribution is marginalized over $7$ nuisance parameters describing the shape of the quasar continuum and coverage corrected according to the procedure described in Section~\ref{sec:inf_test_theory}. Additionally depicted in green is the topology-informed version of the local constraints, entailing the non-trivial prior $P_\mathrm{top}(\logNHI, \Rb)$, explicitly shown in purple in the extra panel, and additionally providing a constraint on the global IGM neutral fraction $\xHI$ in good agreement with the directly inferred one.}
    \label{fig:mock_corner}
\end{figure*}

We start with an illustration of our approach by applying it to the mock spectrum of a quasar which has been shining for $\tQ = 10^{6.25}\,\mathrm{yr}$, observed at the mid-stages of reionization (at $\xHI = 0.55$). The underlying IGM transmission profile is based on a sightline from the semi-numerical reionization topology corresponding to that global neutral fraction value, and we measure $\NHI = 10^{20.25}\,\mathrm{cm}^{-2}$ and $\Rb = 12.9\,\mathrm{cMpc}$ for the local pre-quasar summary statistics of this sightline, corresponding to a typical sightline in an intermediately neutral IGM (c.f. Figure~\ref{fig:P_NHI_Rb_given_xHI}). 
We feed in the same mock spectrum to both our local parameter inference framework introduced in Section~\ref{sec:methods}, and, for comparison, the global version of the pipeline that directly infers the global IGM neutral fraction $\xHI$ rather than the local summaries $\logNHI$ and $\Rb$. Figure~\ref{fig:mock_spec} depicts the mock spectrum with/without noise in black/purple, with the underlying continuum and noise vector shown in green and yellow, respectively. The upper/lower panels depict the reconstructed spectrum (red) and continuum (blue) inferred in the context of the local/global parameterization. While the median inferred models are remarkably similar, the $68\,\%$ scatter (marked by the shaded regions), reflecting parameter uncertainty, continuum reconstruction errors, as well as spectral noise of the inferred model spectrum is significantly smaller in the local parameterization as compared to the global one. This is a direct consequence of the fact that the stochasticity of reionization is an intrinsic part of the model in the global parameterization, but has been removed in the local one as demonstrated in \citet{kist2025b}.

The black contours in the corner plot in
Figure~\ref{fig:mock_corner} depict the coverage corrected local $(\logtQ, \logNHI, \Rb)$ constraints that give rise to the model spectrum shown in the upper panel of Figure~\ref{fig:mock_spec}. Here we already marginalized out the seven nuisance parameters associated to the quasar continuum. We can see that the constraints extend over a vast range in the $(\logNHI, \Rb)$ plane with a mild preference towards higher HI column densities and shorter neutral patch distances. The contours in the $(\logtQ, \logNHI)$ panel hint at a similar degeneracy between these two parameters in the range $10^{19.5}\,\mathrm{cm}^{-2} \lesssim \NHI \lesssim 10^{21}\,\mathrm{cm}^{-2}$ as is often observed between the parameters $\xHI$ and $\logtQ$. 
This degeneracy arises because both a higher HI column density $\logNHI$ (or a higher global IGM neutral fraction $\xHI$), as well as a shorter quasar lifetime $\logtQ$ are causing a stronger IGM damping wing. At $\NHI \lesssim 10^{19.5}\,\mathrm{cm}^{-2}$, however, the axis of degeneracy becomes perfectly vertical. This is because at such low column densities, there ceases to be any identifiable damping wing imprint present in the spectrum, and as such, all constraining power with respect to $\logNHI$ is lost. The size of the proximity zone, on the other hand, places a lower limit on the lifetime of the quasar, since a $\tQ \simeq 10^{3}\,\mathrm{yr}$ quasar can never carve out a proximity zone of this size, regardless of the reionization state of the pre-quasar IGM. The lifetime of this object therefore cannot be significantly smaller than $\tQ \simeq 10^{4}\,\mathrm{yr}$ such that the lifetime constraint extends along the vertical direction. 
A similar effect is observed in the $(\logtQ, \Rb)$ plane where neutral patch distances $\Rb \lesssim 25\,\mathrm{cMpc}$ are preferred at most lifetimes $\tQ \gtrsim 10^{5}\,\mathrm{yr}$, while the contours become vertical at $\tQ \simeq 10^{4}\,\mathrm{yr}$ as our sensitivity to more distant neutral patches vanishes. This is because at lifetimes of $\tQ \gtrsim 10^{5}\,\mathrm{yr}$, a drop in the proximity zone and absorption redward of Lyman-$\alpha$ pushes the constraints toward smaller patch sizes. The reason we can constrain this along with $\logNHI$ is that at fixed $\logNHI$, large values of $\Rb$ would require rare density flucutations at $r > \Rb$, high enough to produce these signatures in the spectrum.

Recall that by assuming the constant prior $P(\logNHI, \Rb)$ that is only limited by the physical boundaries for these parameters, we initially sampled from the topology-\textit{agnostic} posterior $P(\logtQ, \logNHI, \Rb \,|\, \boldsymbol{f})$. Binning these HMC samples into a histogram now allows us to multiply them with the conditional distribution $P_\mathrm{top}(\logNHI, \Rb \,|\, \xHI)$ to fold in the topology dependence via Eq.~(\ref{eq:posterior_final}) and obtain the full four-dimensional topology-informed posterior distribution $P_\mathrm{top}(\xHI, \logtQ, \logNHI, \Rb \,|\, \boldsymbol{f})$ on both global \textit{and} local parameters, shown as green contours in Figure~\ref{fig:mock_corner}. Since the data was not overly constraining with regards to our local summary statistics, the prior $P_\mathrm{top}(\logNHI, \Rb)$---shown explicitly in purple in the extra panel of the same figure---has a clear impact on the shape of the topology-informed posterior which is shifted notably towards higher HI column densities due to the higher prevalence of such sightlines in the realistic ionization topologies (see Figure~\ref{fig:P_xHI_NHI_Rb}). 
The effect on the $\Rb$ posterior is somewhat weaker, but here too we can see that the high-$\Rb$ tail of the distribution gets suppressed. As these low $\logNHI$ and high $\Rb$ values go along with lifetimes of $\tQ \simeq 10^{4}\,\mathrm{yr}$, the long-lifetime mode of the converted posterior obtains significantly more probability mass than in the unconverted case.

Turning to the resulting $\xHI$ constraints, we see that the posterior extends over the entire range of $0 \leq \xHI \leq 1$. In essence, this is because almost any ionization topology contains a number of sightlines with local parameter values similar to the one at hand, as seen in Figure~\ref{fig:P_NHI_Rb_given_xHI}. The probability mass cumulates around intermediate neutral fractions with a peak very close to the true value of $\xHI = 0.55$. Note that the $(\xHI, \logNHI)$ and $(\xHI, \Rb)$ panels of the corner plot can be understood as a reweighted version of the joint
distribution $P_\mathrm{top}(\xHI, \logNHI, \Rb)$ of our local summary statistics shown in Figure~\ref{fig:P_xHI_NHI_Rb}, reweighted according to the posterior distribution $P(\logtQ, \logNHI, \Rb \,|\, \boldsymbol{f})$ as per Eq.~(\ref{eq:posterior_final}).\footnote{As per Eq.~(\ref{eq:joint}), Eq.~(\ref{eq:posterior_final}) is nothing else but the the product of the joint distribution $P_\mathrm{top}(\xHI, \logNHI, \Rb)$ and the topology-agnostic posterior distribution $P(\logtQ, \logNHI, \Rb \,|\, \boldsymbol{f})$.} 
The $(\xHI, \logtQ)$ panel recovers the well-known degeneracy between these two parameters, reflecting the fact that both a higher IGM neutral fraction and a shorter quasar lifetime increase the strength of the damping wing signature. We can now also compare the marginal $(\xHI, \logtQ)$ constraints to the ones obtained by inferring these two parameters directly, without invoking our local parameterization, shown as yellow contours in Figure~\ref{fig:mock_corner}. These constraints are in excellent agreement with the green contours obtained by converting our local constraints.

This demonstrates that our local parameterization allows us to recover the consistent global $\xHI$ constraints on the timing of reionization as conventionally quoted in the literature. However, our local parameterization comes with a number of advantages: first, the global neutral fraction $\xHI$ is not the only parameter constrained by the statistics $\logNHI$ and $\Rb$. Analyzing an ensemble of sightlines in our local parameterization, and determining the distribution of the summaries $\logNHI$ and $\Rb$ provides us with information about the underlying ionization topology, in addition to the constraints on the timing of reionization. 
Second, our local framework facilitates model comparison of different reionization models which can come with different ionization topologies. Their effect can easily be folded into our constraints via the conditional distribution $P_\mathrm{top}(\logNHI, \Rb \,|\, \xHI)$ that provides the stochastic mapping from local to global parameters.
Third, the statistical fidelity of our pipeline with respect to the parameters $\xHI$ and $\logtQ$ improves if we constrain them based on our local parameter constraints rather than infer them directly. We quantify this effect in the subsequent section using the concept of expected coverage probabilities introduced in Section~\ref{sec:inf_test_theory}.

\subsection{Coverage test}
\label{sec:inf_test}

\begin{figure}
	\includegraphics[width=\columnwidth]{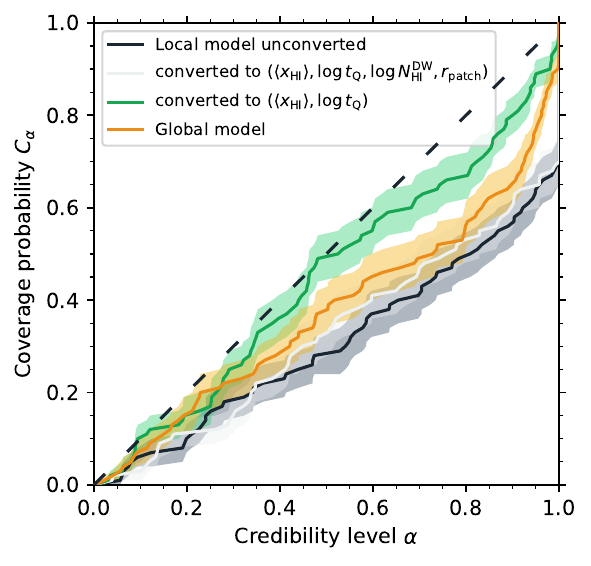}
    \caption{Coverage test results in the form of expected coverage probability $C_\alpha$ as a function of the credibility level $\alpha$ for different sets of model parameters. The coverage is always computed after margnializing over $7$ nuisance parameters describing the shape of the quasar continuum. The coverage probability of the three parameters $(\logtQ, \logNHI, \Rb)$ of our local framework is depicted in black, the coverage after folding in the topology dependence and adding $\xHI$ as an additional parameter in white. The resulting coverage on $\xHI$ and $\logtQ$ after marginalizing over the two local parameters $\logNHI$ and $\Rb$ is depicted in green, and is compared to the coverage curve we obtain when directly inferring $\xHI$ and $\logtQ$ (yellow).}
    \label{fig:inf_tests}
\end{figure}

The mock spectrum discussed in the previous section is part of an ensemble of $10 \times 10$ mock spectra spanning the entire $(\xHI, \logtQ)$ parameter grid (see Section~\ref{sec:mocks}). Each sightline contains an IGM transmission profile drawn from the realistic reionization topology, and hence we can also determine true $\logNHI$ and $\Rb$ values based on the underlying pre-quasar HI density field of each sightline. Having available the ground truth of all four parameters $\xHI$, $\logtQ$, $\logNHI$ and $\Rb$, we can readily perform coverage tests with respect to the unconverted posterior distributions $P(\logtQ, \logNHI, \Rb | \mathrm{\boldsymbol{f}})$, as well as the full four-dimensional converted ones $P_\mathrm{top}(\xHI, \logtQ, \logNHI, \Rb | \mathrm{\boldsymbol{f}})$, and arbitrary marginals thereof.\footnote{Note that before performing any of these coverage tests, we marginalize over all nuisance parameters $\boldsymbol{\eta}$ related to the quasar continuum since we are mainly interested in the coverage on the astrophysical parameters $\boldsymbol{\theta} = (\xHI, \logtQ, \logNHI, \Rb)$.}

We show the results of these coverage tests in Figure~\ref{fig:inf_tests}, where we depict the expected coverage probability $C_\alpha$ as a function of the credibility level $\alpha$, which, in the optimal case, should follow the black dashed one-to-one relation. Firstly, we see that the coverage probability of our local parameter inference (solid black curve) stays consistently below this line, and is therefore overconfident.
This is likely because of the Gaussian approximation of the IGM transmission likelihood (Eq.~\ref{eq:P_t_theta}), which \citet{hennawi2025} identified as the main source of overconfidence in their global parameter inference pipeline. We reproduce their findings by performing a second coverage test on the directly inferred parameters $\xHI$ and $\logtQ$, based on the same ensemble of mock spectra. The overconfidence of this coverage curve, depicted in yellow in Figure~\ref{fig:inf_tests} \citep[see also the left-hand panel of Figure~11 in][]{hennawi2025}, is somewhat more mild than in the local parameterization (black).
The reason for this is likely that the local parameterization has removed a notable degree of scatter from the IGM transmission profiles. While this additional amount of scatter acts as a major additional source of uncertainty for the final parameter constraints \citep{kist2025a}, it also aids the inference by gaussianizing the likelihood function according to the central limit theorem. 
Our Gaussian approximation to the IGM transmission likelihood (which applies in both cases) is therefore slightly better justified in the global parameterization than in the local one, leading to the somewhat better (but still suboptimal) coverage behavior.

The coverage of the four-dimensional, converted posterior distribution $P(\xHI, \logtQ, \logNHI, \Rb | \mathrm{\boldsymbol{f}})$ (white curve) looks almost identical to the unconverted, three-dimensional one since it remains affected by the overconfidence with respect to the local parameters $\logNHI$ and $\Rb$. On the other hand, after marginalizing out these two parameters, we are left with a nearly perfect coverage curve on the remaining two parameters $\xHI$ and $\logtQ$ (green curve). Notably, the coverage behavior is even better than if we had inferred these two parameters directly. This shows that our local parameter inference pipeline can be advantageous even if all interest is geared towards global constraints on the timing of reionization, disregarding the local summaries $\logNHI$ and $\Rb$ as nuisance parameters. This result is somewhat surprising given the overconfidence when inferring the local parameters on their own. However, it is important to realize that the coverage on other parameters, subsequently obtained from those, is not necessarily similarly bad.

In the case at hand, note that the mapping from local to global parameters is probabilistic, and therefore in particular highly non-injective, i.e., many local parameter combinations can get mapped to the same global IGM neutral fraction $\xHI$ (c.f. Figure~\ref{fig:P_NHI_Rb_given_xHI}). Even if there are imperfections in the local $\logNHI$ and $\Rb$ constraints, these parameters can still get mapped to the 'correct' global $\xHI$ value. Specifically, this appears to be the case here to such a high degree that the converted $(\xHI, \logtQ)$ constraints are statistically more faithful than the directly inferred ones. An additional factor that likely contributes is that binning the conditional distribution $P(\logNHI, \Rb | \xHI)$ in order to convert the constraints via Eq.~(\ref{eq:posterior_final}) smooths out the contours and thus improves the coverage to a certain extent,
whereas such a binning step is not required when inferring $\xHI$ and $\logtQ$ directly.%

\subsection{Inference precision}
\label{sec:precision}

\begin{figure*}
	\includegraphics[width=\textwidth]{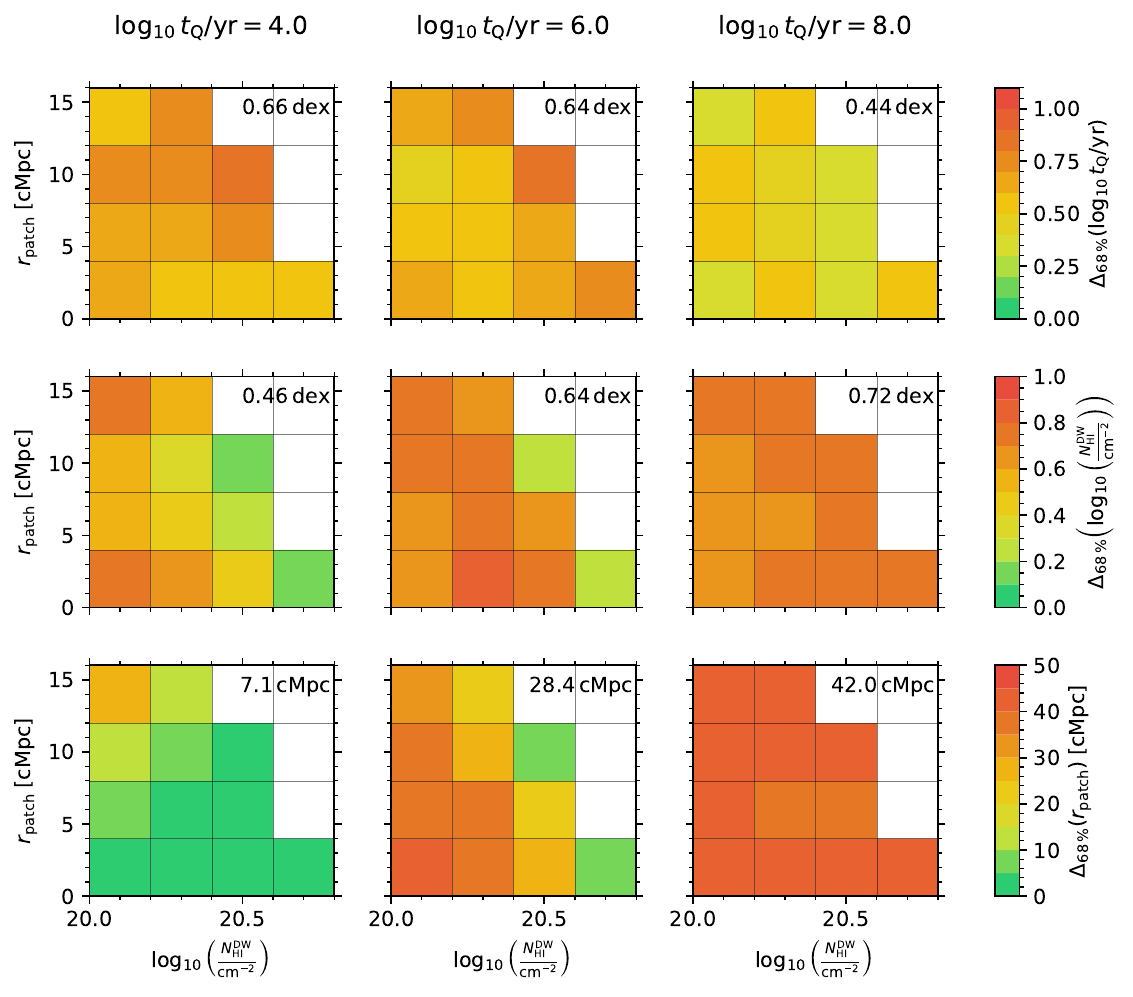}
    \caption{Inference precision with respect to the three parameters of our local damping wing parameterization in different regions of parameter space. Precision on $\logtQ$, $\logNHI$ and $\Rb$ is shown in the top, middle and bottom row, respectively. The lifetime is fixed to $\tQ = 10^4$, $10^6$, and $10^8\,\mathrm{yr}$ in the left, middle and right column, respectively, and each panel depicts the precision in the region of $(\logNHI, \Rb)$ parameter space where damping wings are present. Each pixel shows the average precision values of $10$ different mock objects from mock ensemble (ii) described in Section~\ref{sec:mocks}, where the same $10$ continua are used at each location in parameter space to isolate the effects of the astrophysical parameter variations. Annotated values are the mean of all precision values in a given panel.}
    \label{fig:inf_prec}
\end{figure*}

\begin{figure*}
	\includegraphics[width=\textwidth]{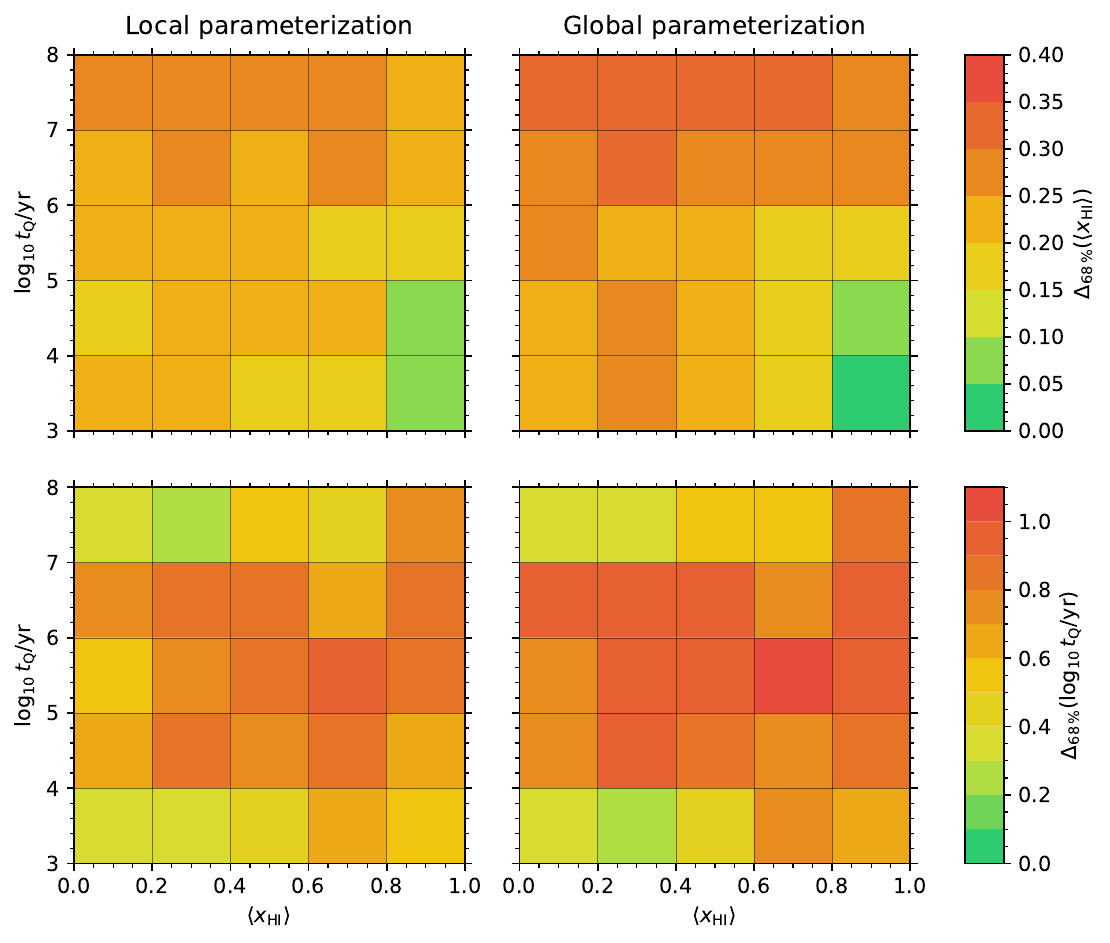}
    \caption{Inference precision with respect to the two parameters $\xHI$ and $\logtQ$ of the global damping wing parameterization in different regions of parameter space, compared between our local and global inference framework. Precision on $\xHI$ is shown in the upper row, precision on $\logtQ$ in the lower row. The left column shows precision values of constraints converted from the local to the global parameterization, whereas the ones in the right column correspond to $(\xHI, \logtQ)$ inferred directly. Each pixel shows the average precision values of $4$ different mock objects originating from realistic ionization topologies, i.e., based on the mock ensemble (i) described in Section~\ref{sec:mocks}.}
    \label{fig:inf_prec_global_vs_local}
\end{figure*}

Having investigated the coverage behavior of our inference framework, we now proceed by quantifying its precision. We do so by considering the one-dimensional marginal posterior distribution of a given astrophysical parameter, and computing its width
\begin{equation}
\label{eq:prec}
    \Delta_{68\,\%}(\theta) = \frac{1}{2} (p_{84\,\%}(\theta) - p_{16\,\%}(\theta))
\end{equation}
as a measure of the $1\sigma$ error on the parameter $\theta\in\{\xHI, \logtQ, \logNHI, \Rb\}$, where $p_{q}(\theta)$ is the $q$-th percentile of the respective marginal posterior distribution. Before determining these percentiles, we perform a coverage test and reweigh the inferred posterior distribution according to the prescription discussed in Section~\ref{sec:inf_test_theory} and \citet{hennawi2025} to ensure that we do not quote overly optimistic precision values.

\subsubsection{Local parameterization}

We start by investigating the precision we can achieve on $\logtQ$, $\logNHI$ and $\Rb$ when performing the inference in our new local parameterization. Here we only determine the inference precision in the region of parameter space where a non-negligible damping wing imprint is present, i.e., where we can expect at least some degree of sensitivity to these parameters. In other words, we choose to concentrate our computational resources for this exercise to regions where our constraints are not fully prior-dominated.
These are sightlines with comparably high HI column densities $\logNHI$ and shorter distances $\Rb$ to the first neutral patch. We therefore consider a $4 \times 4$ grid in $(\logNHI, \Rb)$ parameter space with values of $20.1 \leq \logNHI  \leq 20.7$ and $2\,\mathrm{cMpc} \leq \Rb \leq 14\,\mathrm{cMpc}$, i.e., mock ensemble (ii) described in Section~\ref{sec:mocks}. In what follows, the configuration with $\logNHI = 20.1$ and $\Rb = 14\,\mathrm{cMpc}$ can be seen as representative of the remaining parts of parameter space where $\logNHI < 20.1$ and/or $\Rb > 14\,\mathrm{cMpc}$. Note that due to the physical constraints on the possible $(\logNHI, \Rb)$ combinations, models only exist for $12$ out of the $16$ points on this grid. Recall that it does \textit{not} take a topology-informed prior $P_\mathrm{top}(\logNHI, \Rb)$ to exclude the remaining $4$ parameter combinations. These are \textit{physically excluded} based on the topology-independent distribution of density fluctuations in the IGM.

We add a third grid dimension by considering three different lifetime values of $\tQ = 10^4, 10^6$ and $10^8\,\mathrm{yr}$. At each location in $(\logtQ, \logNHI, \Rb)$ parameter space, we consider $10$ distinct mock spectra, and we average the inferred precision values over these $10$ objects. To further suppress variations throughout parameter space due to other sources of stochasticity, we consider the \textit{same} $10$ continuum draws and noise realizations at each location in parameter space. Overall, this amounts to a set of $12 \times 3 \times 10 = 360$ mock spectra. Note that due to the fact that these parameter values 
do not span 
the full prior range of the three astrophysical parameters $(\logtQ, \logNHI, \Rb)$, we do not determine the coverage behavior on this set of mocks but rather on the one considered in Section~\ref{sec:inf_test}. This is possible since all that is required for the reweighting procedure outlined in Section~\ref{sec:inf_test_theory} is the functional relationship between credibility levels $\alpha$ and expected coverage probability $C_\alpha$. Practically, this means we consider the unconverted 
local 
parameter coverage (black curve) from Figure~\ref{fig:inf_tests}, and use it to reweigh our $(\logtQ, \logNHI, \Rb)$ samples obtained from the new mock ensemble introduced in this section. We then compute the inference precision based on the percentiles of the \textit{reweighted} samples according to Eq.~(\ref{eq:prec}).

The averages of these precision values as a function of the astrophysical parameter values are plotted in Figure~\ref{fig:inf_prec} where we show from top to bottom the inference precision on $\logtQ$, $\logNHI$ and $\Rb$, respectively. The three columns represent the three different mock quasar lifetimes of $\tQ = 10^4, 10^6$ and $10^8\,\mathrm{yr}$, respectively, and each panel itself depicts the precision in $(\logNHI, \Rb)$ parameter space. The annotated values are the mean precision of all mocks in the respective panel.

We begin by analyzing the lifetime precision $\Delta_{68\,\%}(\logtQ)$ for which we do not identify a notable dependence on the local parameters $\logNHI$ and $\Rb$. The mean precision values are $0.66\,\mathrm{dex}$ at $\tQ = 10^4\,\mathrm{yr}$ and $0.64\,\mathrm{dex}$ at $\tQ = 10^6\,\mathrm{yr}$, decreasing to $0.44\,\mathrm{dex}$ at $\tQ = 10^8\,\mathrm{yr}$, as seen in the upper row of Figure~\ref{fig:inf_prec}. This enhanced precision at long lifetimes is in line with the trend identified in \citet{kist2025a} in the context of the global parameterization, attributed to the thermal proximity effect due to Helium II ionization. This becomes apparent for quasars with long lifetimes and thus a significantly extended Helium ionization front where the associated thermal heating has enhanced the strength of the IGM transmission peaks, making such lifetimes easier to identify through this distinct feature. We refrain from a quantitative comparison of our lifetime precision values to those quoted in \citet{kist2025a} due to the fact that due to the different parameterizations, an apples-to-apples comparison of the same regions of parameter space is challenging, and, more importantly, we are investigating here the precision of our \textit{topology-agnostic} constraints, whereas the global constraints in \citet{kist2025a} are inevitably topology-informed. Instead, we proceed in the subsequent section by performing a comparison of the precision values we can obtain in global $(\xHI, \logtQ)$ parameter space after conversion of our local constraints versus directly inferring $\xHI$ and $\logtQ$.

For now, proceeding to the HI column density inference precision $\Delta_{68\,\%}(\logNHI)$, we observe similar trends as found by \citet{kist2025a} for the global IGM neutral fraction $\xHI$, that is, an increasingly high precision the stronger the IGM damping wing. This is unsurprising given that the strongest imprints can most easily be disentangled from the intrinsic continuum of the quasar. This trend is clearly seen when focusing on rows of fixed $\Rb$ at short to intermediate quasar lifetimes in the first two panels of the middle row of Figure~\ref{fig:inf_prec}. Vice versa, at fixed $\logNHI$ (i.e., in a given column of one these panels), the inference precision on this parameter appears to deteriorate the closer the location $\Rb$ of the first neutral patch to the quasar. A possible reason for this is the somewhat increased scatter of IGM transmission profiles in the local parameterization at small $\Rb$ values \citep[see Figures~8 and 9 in][]{kist2025b} which makes it harder to reconstruct the true column density of a given sightline.

Both aforementioned trends with $\logNHI$ and $\Rb$ cease to exist when the lifetime of the quasar is long ($\tQ = 10^8\,\mathrm{yr}$) where we find a uniform $\logNHI$ inference precision across the entire range of parameter space considered in this figure. This is likely because the damping wing imprint is so weak at these lifetimes that it can hardly be disentangled from the intrinsic spectrum of the quasar, even at the highest HI column densities $\logNHI$.

The most significant trends are seen for the inference precision on the distance $\Rb$ of the quasar to the first neutral patch, depicted in the bottom row of Figure~\ref{fig:inf_prec}. At the shortest lifetime of $\tQ = 10^4\,\mathrm{yr}$, we can identify a highly pronounced trend of enhanced precision at higher HI column densities $\logNHI$ and shorter neutral patch distances $\Rb$, improving from $\sim 40\,\mathrm{cMpc}$ down to $<5\,\mathrm{cMpc}$. These high precision values can be explained by the fact that the location of the ionization front for low-lifetime objects coincides well with the pre-quasar neutral patch location since such quasars have not yet carved out an ionized patch that extends significantly beyond this location. As a result, $\Rb$ can easily be reconstructed from the spectrum of such objects. However, if the patch location is too far away, the decrease of the quasar's photoionization rate $\Gamma_\mathrm{QSO}$ according to the inverse-square law, $\Gamma_\mathrm{QSO} \sim 1/r^2$, suppresses all transmission already at a distance closer to the quasar than $\Rb$, and hence we lose sensitivity to this parameter \citep[see also Figure~8 in][]{kist2025b}.

At longer quasar lifetimes, inference precision on $\Rb$ deteriorates since most such objects have ionized away the first pre-quasar neutral patch, making it significantly harder (if not even entirely impossible) to reconstruct its location. At $\tQ = 10^6\,\mathrm{yr}$, a relatively high precision is retained for the $(\NHI, \Rb) = (10^{20.7}\,\mathrm{cm}^{-2}, 2\,\mathrm{cMpc})$ and $(10^{20.5}\,\mathrm{cm}^{-2}, 10\,\mathrm{cMpc})$ models. These are parameter combinations where the location of the ionization front after $10^6\,\mathrm{yr}$ still coincides well with the location of the first pre-quasar neutral patch \citep[compare also Figure~9 in][]{kist2025b}, and as such, is still reconstructible at a reasonable precision. On the other hand, after shining for $\tQ = 10^8\,\mathrm{yr}$, the quasar has ionized away so much neutral material that virtually any information about the location of the pre-quasar neutral patch is lost, regardless of the location in $(\logNHI, \Rb)$ parameter space.

In summary, we have found that in certain regions of parameter space, we are sensitive to all three parameters of our local IGM damping wing parameterization, demonstrating its value for astrophysical parameter inference. Specifically, we saw that we can reconstruct the quasar lifetime $\logtQ$ to $\sim 0.4 - 0.7 \,\mathrm{dex}$, the HI column density to $\sim 0.7 \,\mathrm{dex}$, improving down to $\sim 0.2 \,\mathrm{dex}$ when the damping wing is strong, and the neutral patch distance $\Rb$ up to $<5\,\mathrm{cMpc}$ at short quasar lifetimes of $\tQ \simeq 10^4\,\mathrm{yr}$, whereas sensitivity gets lost the longer the quasar has been shining.

\subsubsection{Global parameterization}

As pointed out in the previous section, an apples-to-apples comparison of the inference precision achieved in the local parameterization to that quoted in \citet{kist2025a} in the context of the conventional global parameterization is not possible. To perform such a comparison, we instead consider mock ensemble (i) described in Section~\ref{sec:mocks} where the $x_\mathrm{HI}$ sightlines originate from realistic semi-numerical ionization topologies and we can compare the precision achieved on $\xHI$ and $\logtQ$ by converting the local $(\logtQ, \logNHI, \Rb)$ constraints to these global parameters versus inferring them directly.

Figure~\ref{fig:inf_prec_global_vs_local} depicts these precision values obtained on $\xHI$ in the upper row and on $\logtQ$ in the lower row, comparing the converted local constraints in the left column to the directly inferred global ones in the right one. Each panel by itself shows the precision values in the $(\xHI, \logtQ)$ parameter plane, where each pixel corresponds to the average precision of $4$ mocks from our $10 \times 10$ mock object ensemble. Note that there is a direct correspondence between the two right-hand panels and the bottom-right panels of Figures~9 and 10 in \citet{kist2025a}. We average the precision over $4$ mocks in each pixels to suppress sightline-to-sightline stochasticity as far as possible. Note that to that end, our mock ensemble considered in the previous section had fixed continua and noise draws across the entire range of parameter space. This is \textit{not} the case in this section as we are only interested in the differences in inference precision based on the two parameterizations, so a one-to-one comparison between fixed pixels in the left and right panels is still possible.

What we find when performing this comparison is a near perfect agreement between the precision values obtained in the two parameterizations. Small differences only become apparent in $\xHI$ precision towards the (high-$\xHI$, low-$\logtQ$) end where the global constraints become somewhat more precise. This is a consequence of the binning we adopt when converting the local to the global constraints which are limited by the $\xHI$ grid spacing of $0.05$. On the other hand, the converted constraints are slightly more precise in regions of low precision (e.g. $\xHI$ precision at the long-lifetime end $\tQ \simeq 10^8\,\mathrm{yr}$, or $\logtQ$ precision at intermediate lifetimes $\tQ \simeq 10^{5-6}\,\mathrm{yr}$). This likely results from the stronger coverage corrections that have to be applied in the context of the global parameterization where the inference is somewhat overconfident.

Besides these minor differences, our main conclusion is that the inference precision on $\xHI$ and $\logtQ$ coincides remarkably well, regardless of the parameterization that the inference was originally performed in. This demonstrates that we are not losing any information on the global timing of reionization by first inferring local constraints and subsequently tying them to a global reionization model. Note that on the other hand, we also cannot expect significantly tighter constraints than from the direct global parameter inference since the conversion step necessarily implies that we are again folding in the stochasticity of reionization that our local parameterization has removed from the inference task itself. However, our new parameterization allows us to gain hitherto unused information about the local ionization topology in front of a quasar while still retaining the same precision on the parameters $\xHI$ and $\logtQ$ when inferring them directly in the context of the global parameterization.

\section{Conclusions}
\label{sec:conclusions}

IGM damping wings towards quasars are a unique probe not only of the history of reionization but also its topology, and \citet{kist2025b} introduced a new set of local summary statistics encapsulating all this information. Here we presented a fully Bayesian inference framework that allows us to extract this information in a topology-independent fashion. These constraints are informative about the local ionization topology before the quasar started shining, and, when tied to a specific reionization model, also about the global timing of reionization.

To establish this connection, we introduced a probabilistic framework that allows us to map our local damping wing statistics, consisting of a Lorentzian-weighted HI column density
$\logNHI$ as well as the distance $\Rb$ 
of the quasar to the first neutral patch in the pre-quasar topology, to the global volume-averaged IGM neutral fraction $\xHI$. The virtue of our approach is that the two local summaries $\logNHI$ and $\Rb$, along with the lifetime $\logtQ$ of the quasar, can be inferred independently of any assumptions about the reionization model. We demonstrated that all these assumptions can be encoded in a non-trivial prior $P_\mathrm{top}(\logNHI, \Rb)$ which can be folded in \textit{subsequently}, facilitating the comparison of different reionization models.

In addition, we found that, after marginalizing over the local constraints, this two-step procedure improves the statistical fidelity of our inference pipeline. The inference of the local damping wing statistics themselves shows a certain degree of overconfidence, but we correct for this by broadening the inferred posteriors in a principled way before quoting actual constraints. In the future, simulation-based inference approaches will be key to also capture non-Gaussianities in the IGM transmission likelihood, and hence make use of the full information contained in the spectra while removing the need for coverage correction altogether \citep{chen2024}.

Based on a large sample of mock spectra, we quantified that precision at which we can infer the three astrophysical model parameters, and we found that we can constrain the quasar lifetime to $0.58_{-0.13}^{+0.13}\,\mathrm{dex}$, the HI column density to $0.69_{-0.30}^{+0.06}\,\mathrm{dex}$, and the distance to the first neutral patch to $31.4_{-28.1}^{+10.7}\,\mathrm{cMpc}$ for model parameter combinations that would imprint a significant damping wing upon the spectrum of the quasar.
Furthermore, we found that after tying our local constraints to a given reionization model, our precision on the global IGM neutral fraction $\xHI$ and the quasar lifetime $\logtQ$ is on par with the precision achieved when inferring these parameters directly in the context of the conventional, global parameterization, while removing the need for coverage corrections. This advantage comes at the price of a somewhat increased computational cost due to the increased dimensionality of parameter space which also increases the number of required simulation models.

The biggest virtue of adopting our approach, though, is that it provides model-independent, physical information about the local ionization topology in front of a given quasar. If concerned with multiple sightlines at a similar redshift, setting up a Bayesian hierarchical model will allow us to combine these topology-agnostic local constraints to a topology-informed constraint not only on the global timing of reionization but also its topology \citep{sharma2025}. In addition, the local information we gain through these statistics will allow us to draw novel connections to the 21 cm power spectrum, and the ionized bubble sizes measured in studies of Lyman-$\alpha$ emission from galaxies. We will further explore these connections in future work.

\section*{Acknowledgements}

We acknowledge helpful conversations with the ENIGMA group at UC Santa Barbara and Leiden University, especially with Da-Ming Yang about the low-redshift continuum data. This work made use of \texttt{NumPy} \citep{harris2020}, \texttt{SciPy} \citep{virtanen2020}, \texttt{JAX} \citep{bradbury2018}, \texttt{NumPyro} \citep{bingham2018, phan2019}, \texttt{sklearn} \citep{pedregosa2011}, \texttt{Astropy} \citep{astropy_collaboration2013, astropy_collaboration2018, astropy_collaboration2022}, \texttt{PypeIt} \citep{prochaska2020}, \texttt{SkyCalc\_ipy} \citep{leschinski2021}, \texttt{h5py} \citep{collette2013}, \texttt{Matplotlib} \citep{hunter2007}, \texttt{corner.py} \citep{foreman-mackey2016}, and \texttt{IPython} \citep{perez2007}.
TK and JFH acknowledge support from the European Research Council (ERC) under the European Union’s Horizon 2020 research and innovation program (grant agreement No 885301). JFH acknowledges support from NSF grant No. 2307180.

\section*{Data Availability}

The derived data generated in this research will be shared on reasonable requests to the corresponding author.

\bibliographystyle{mnras}
\bibliography{main} %

\appendix

\bsp	%
\label{lastpage}
\end{document}